\documentclass[twocolumn]{aastex62}
\usepackage{textcomp}
\usepackage{amssymb}
\usepackage{wasysym}
\usepackage{CJK}
\usepackage{enumerate}

\submitjournal{ApJ}

\shorttitle{CIRCUMNUCLEAR X-RAY EMISSION OF NGC 2992}
\shortauthors{Xu \& Wang}

\begin{document}
\begin{CJK*}{UTF8}{gbsn}
\title{Revisit the Circumnuclear X-ray Emission of NGC 2992 in a Historically Low State}

\correspondingauthor{Junfeng Wang}
\email{jfwang@xmu.edu.cn}

\author[0000-0003-0970-535X]{Xiaoyu Xu (许啸宇)}
\affil{Department of Astronomy, Xiamen University, Xiamen, Fujian 361005, China}

\author[0000-0003-4874-0369]{Junfeng Wang}
\affiliation{Department of Astronomy, Xiamen University, Xiamen, Fujian 361005, China}

\collaboration{}

\begin{abstract}

The inner-most region of the Seyfert galaxy NGC 2992 has long been suspected to be the location of intense AGN-host galaxy interaction, but photon pile-up in previous high-resolution observations hampered the study of soft X-ray excess and the interaction near its nucleus. 
We present an X-ray imaging spectroscopic analysis of the circumnuclear ($1\arcsec$--$3\arcsec$) region of NGC 2992 using the zeroth-order image of a 135 ks grating observation obtained with {\em Chandra}, which captured the nucleus in a historically low flux state.  Extended soft X-ray emission is detected in the circumnuclear region with observed luminosity $L_{\rm X} \sim 7 \times 10^{39}\rm\ erg\ s^{-1}$. The majority of previously puzzling detection of soft excess could be associated with the outflow, indicated by the morphological correspondences between soft X-ray emission and figure-eight-shaped radio bubbles. 
An anomalous narrow emission line with the centroid energy $\sim4.97$ keV is found. 
If attributed to redshifted highly ionized iron emission (e.g., Fe~{\sc{xxv}}), the required outflow velocity is $\sim0.23\,c$. 
An alternative explanation is that this line emission could be produced by the nuclear spallation of iron.  We also find asymmetric extended Fe K$\alpha$ emission along the galactic disk, which could originate from reflection by cold gas on $\sim 200$ pc scale.

\end{abstract}


\section{Introduction} \label{sec:intro}

It is generally expected that actively accreting supermassive black holes (SMBHs) as active galactic nuclei (AGNs) can have a profound impact upon the environment and evolution of their host galaxies, releasing energy in radiation and powerful outflow \citep{2012ARA&A..50..455F,2015ARA&A..53..115K}.
The AGN-driven outflows may take forms of highly collimated jets or winds with a large opening angle when interacting with the circumnuclear interstellar medium (ISM) and galaxy-scale gas.
Evidence for outflows in neutral (both atomic and molecular) and ionized phases has been vastly observed \citep[e.g.,][]{2008A&A...491..407N,2010A&A...518L.155F,2011ApJ...742...44T,2012MNRAS.426.1073H,2013ApJ...776...27V,2014ApJ...788...54G,2014MNRAS.441.3306H, 2015A&A...578A..11B}.
Yet the exact physical processes of AGN feedback and the quantitative impact on its host galaxy remains an open question.

Spatially resolved multi-wavelength studies of circumnuclear regions on the kpc scale become powerful in tackling this issue over the years \citep[e.g.,][]{2017MNRAS.467.2612W,2018A&A...619A..74V}.
In the X-rays, extended narrow emission-line regions in local Seyfert galaxies have been demonstrated to be useful probes of AGN driven outflow, jet-ISM collision \citep[e.g.,][]{2012ApJ...756..180W}, starburst \citep[e.g.,][]{1999ApJ...523..575L,2004ApJ...606..829S,2009ApJ...704.1195W,2020ApJ...903...35H}, and/or AGN-photoionization \citep[e.g.,][]{2011ApJ...742...23W,2014ApJ...781...55W,2018ApJ...855..131F,2022arXiv220213803F}.
In this work, we describe our findings from X-ray observations of NGC 2992, with a complex new structure surrounding the AGN unveiled.

NGC 2992 is a nearly edge-on spiral galaxy at $z=0.00771$ \citep{1996ApJS..106...27K} which corresponds to a distance of 32.5 Mpc (1\textacutedbl$\sim150$ pc) assuming a Hubble constant $H_{0}=75\rm\ km\,s^{-1}$, and its classification is Seyfert 1.9/1.5 \citep{2008AJ....135.2048T}. 
The first X-ray detection of NGC 2992 is in 1977 by HEAO-1 at a flux level of about $\rm 8\times10^{-11}\ erg\,cm^{-2}\,s^{-1}$ \citep{1982ApJ...253..485P}. 
It has long been known that the 2-10 keV flux of NGC 2992 can vary by a factor of $\sim30$ with the lowest 2-10 keV flux ($3\times 10^{-12}\rm\,erg\,cm^{-2}\,s^{-1}$) was observed in 2010 by {\em XMM-Newton} and the highest ($1.1\times 10^{-10}\rm\,erg\,cm^{-2}\,s^{-1}$) in 2019 by {\em Swift-XRT} \citep{2007ApJ...666...96M,2018MNRAS.478.5638M,2022MNRAS.514.2974M}. 
In 2010, the source was also captured by the {\em Chandra X-ray Observatory} in the historically low state, with a 2--10 keV flux of $\rm 3.6\times10^{-12}\,erg\,cm^{-2}\,s^{-1}$ \citep{2017ApJ...840..120M}, allowing for study of the nucleus with the highest spatial resolution in the X-rays.  

Spatially extended soft X-ray and significant soft excess in the spectrum have been observed in NGC 2992 previously by {\em ROSAT} \citep{1998ApJ...496..786C}. 
\cite{2005ApJ...628..113C} have studied this source in detail and found several extranuclear X-ray sources with distances $r \gtrsim 3\arcsec$ from the nucleus by {\em Chandra}.  
However, the flux of these sources cannot account for most of the soft excess.  Thus, they suggested that the majority of the soft excess emission is originated from a region between $1\arcsec$ and $3\arcsec$ from the nucleus, which is not imaged in their observation due to the severe pileup effect. 
They noted that the soft X-ray emission of NGC 2992 could be produced either by a hot wind driven by starburst or AGN, or by AGN photoionization and photoexcitation. In fact, \cite{2007MNRAS.374.1290G} conclude that the soft excess is dominated by photoionization of circumnuclear gas, using the {\em XMM-Newton} RGS data spectrum of NGC 2992, with detection of C~{\sc{vi}} Ly$\rm \beta$, O~{\sc{vii}} He$\rm \alpha$(r), O~{\sc{viii}} Ly$\rm \alpha$ and O~{\sc{viii}} Ly$\rm \beta$ \citep[see also][for discussion on the weak signal to noise ratio]{2010ApJ...713.1256S}.

NGC 2992 has two prominent diffuse figure-eight-shaped radio bubbles to the northwest and southeast, which have a size of about 8\arcsec with the axis misaligned from the major axis of the galaxy by about $26^{\circ}$ \citep{1984ApJ...285..439U}. 
\cite{2000MNRAS.314..263C} proposed that the radio bubbles were driven by AGN with their near-infrared observations.  
\cite{2001A&A...378..787G} further suggested that the figure-eight-shaped radio structure could well be expanding bubbles according to the spatial correlation between an extended arc-shaped emission detected in [O~{\sc{iii}}] and the radio bubbles.  Because the X-ray peak is coincident with the core of the bubbles, \cite{2005ApJ...628..113C} suggested that the radio feature could be inflated by an AGN instead of a starburst. More recently, \citet{2017MNRAS.464.1333I} reported on C-band (5--7 GHz) observation of NGC 2992 and a clear detection of nuclear bipolar radio outflow revealed in linearly polarized emission, strongly support presence of the AGN-related outflow. The bend in the west lobe indicates that the radio outflows are interacting with the ISM. The aforementioned radio bubble is at smaller scale, referred to as limb brightened ``double loop'', was not resolved. 

A biconical outflow along the minor axis of the galactic disk has been detected in the optical band \citep[e.g.][]{2001AJ....121..198V,2019A&A...622A.146M}. 
The wind extends to $\sim$2.8 kpc aligned with the large-scale radio structure \citep[][]{1996ApJ...467..551C} but misaligned by $\sim44^{\circ}$ with figure-eight radio bubbles, with velocities of the outflow range from 50 to 200 $\rm km\,s^{-1}$. 
\cite{2001AJ....121..198V} concluded that the outflow was most likely an AGN-driven wind according to the diffuse radio morphology and absence of direct evidence of powerful nuclear starburst in NGC 2992. 
Using the adaptive optics near-infrared integral field spectrograph SINFONI on the Very Large Telescope (VLT), \cite{2010A&A...519A..79F} suggested that the large-scale outflow and figure-eight-shaped radio bubbles were driven by the AGN rather than the starburst, considering the relevant energy budget and dynamical timescale. 
Recently, the primary ionization source of the outflow was confirmed to be the AGN based on the spatially resolved Baldwin, Phillips, and Telervich \citep[BPT,][]{1981PASP...93....5B} diagnostic map \citep[][]{2019A&A...622A.146M,2021MNRAS.502.3618G}.
Two redshifted and ionized Si emission lines were observed in the X-ray band, with the High Energy Transmission Grating Spectrometer \citep[HETGS;][]{2005PASP..117.1144C} onboard {\em Chandra}, which indicate presence of a photoionized fast outflow with a velocity of $\sim$2500 $\rm km\ s^{-1}$ \citep{2017ApJ...840..120M}.  
\cite{2018MNRAS.478.5638M} also found possible evidence of an ultra-fast outflow (UFO) in NGC 2992 with velocity $v_{\rm out}\sim$0.21$c$, where $c$ is the speed of light, which was measured using an absorption line which could be identified as Fe~{\sc{xxv}} captured by {\em XMM-Newton} and {\em NuSTAR} in a medium-high flux state. 

In this paper, we present spectra and spatial images of a region around the nucleus ($r\sim 1\arcsec-3\arcsec$) obtained from a $\sim$135 ks {\em Chandra} observation of NGC 2992 in 2010, to investigate the previously inaccessible X-ray emission near the nucleus. 
Due to the historically low flux state of NGC 2992 and the low count rate of zeroth-order images with HETGS, we can overcome the severe pile-up effect and resolve the circumnuclear region. 
In Section 2, we describe the observations and reduction of the data. We present our results on the extended soft X-rays and line emission in the hard X-ray band in Sections 3 and 4, respectively.  We discuss these findings in Section 5, and finally summarize in Section 6.

\section{Observations and data reduction} \label{sec:data}

During 2010, NGC 2992 was observed three times (ObsIds 11858, 12103, and 12104), with exposure times 94.4 ks, 19.2 ks, and 21.9 ks respectively, by the HETG in combination with the Advanced CCD Imaging Spectrometer-Spectroscopy \citep[ACIS-S;][]{2003SPIE.4851...28G} array onboard {\em Chandra} \citep[see][for details of observations]{2017ApJ...840..120M}. Previous work by \cite{2017ApJ...840..120M} have analyzed these data thoroughly to obtain the first-order High Energy Grating (HEG) and Medium Energy Grating (MEG) spectra.

CIAO version 4.9 and CALDB version 4.7.6 were used to reprocess the data. 
We used the reprocessing script {\tt chandra\_repro} to create new ``level 2'' event files, then combined these files into a deep exposure by the script {\tt merge\_obs}. From the new merged event file, we obtained zeroth-order imaging data with a total exposure time of $\sim$ 135 ks. The X-ray spectra of the circumnuclear region (1\arcsec-3\arcsec) were extracted from event files by using the CIAO script {\tt specextract}. The background spectrum was obtained from a circular region (the radius $r=30 \arcsec$) in the same CCD chip free of point sources. 
The spectral fitting was performed with the XSPEC package \citep[version 12.9.1m][]{1996ASPC..101...17A}.

\section{Resolving the Soft X-ray Excess} \label{sec:soft}
\subsection{Spatial analysis} \label{subsec:soft}
\cite{2005ApJ...628..113C} have previously noted extended asymmetric soft X-ray emission from $\sim$2.5\arcsec\ to 4\arcsec\ in 0.3--1.0 keV band using the {\em Chandra} ACIS data without grating. However, due to pile-up, they could not resolve the emission within $\sim$2.5\arcsec.

\begin{figure*}[htbp!]
	\gridline{\fig{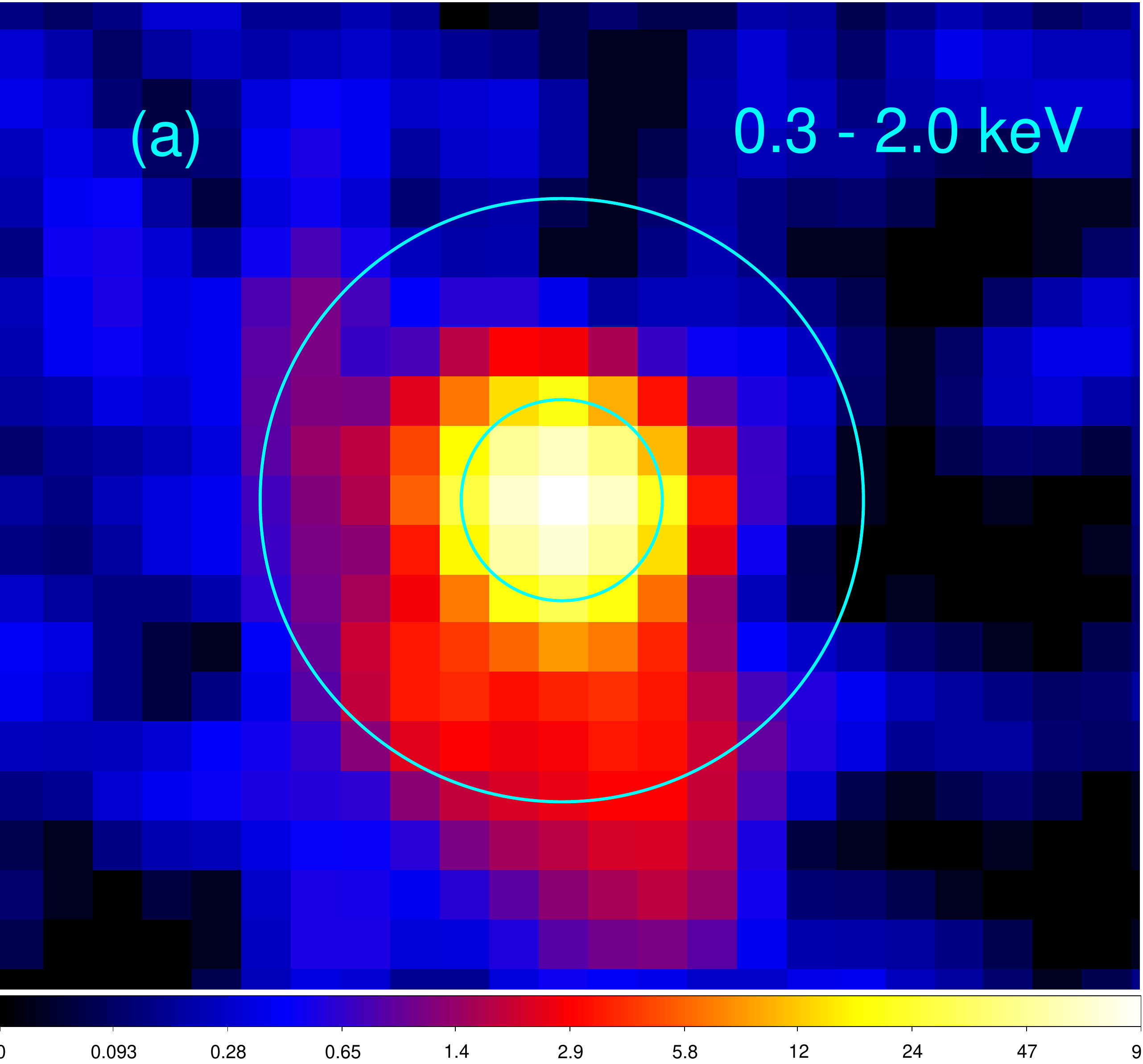}{0.3\textwidth}{}
		\fig{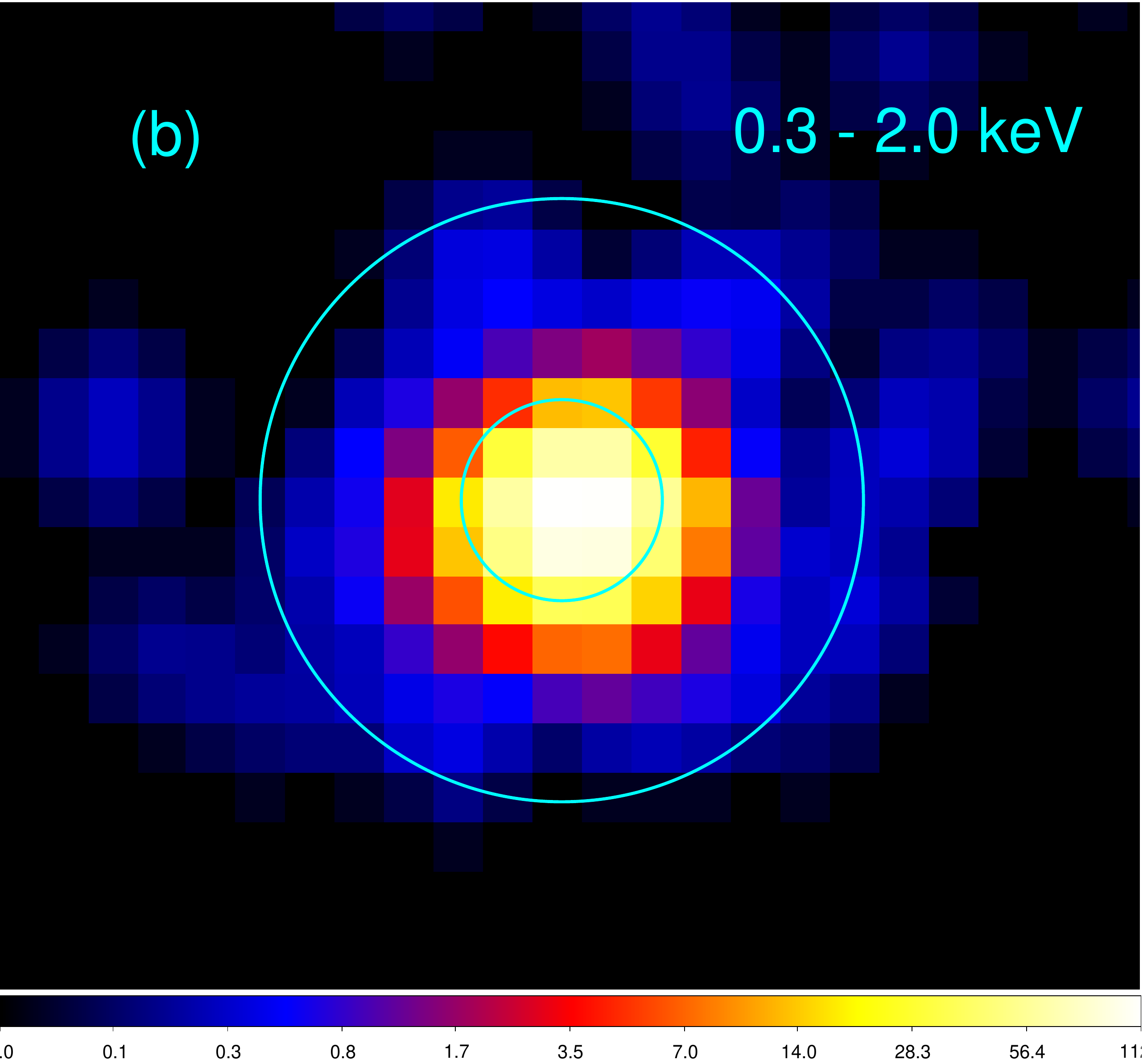}{0.3\textwidth}{}
		\fig{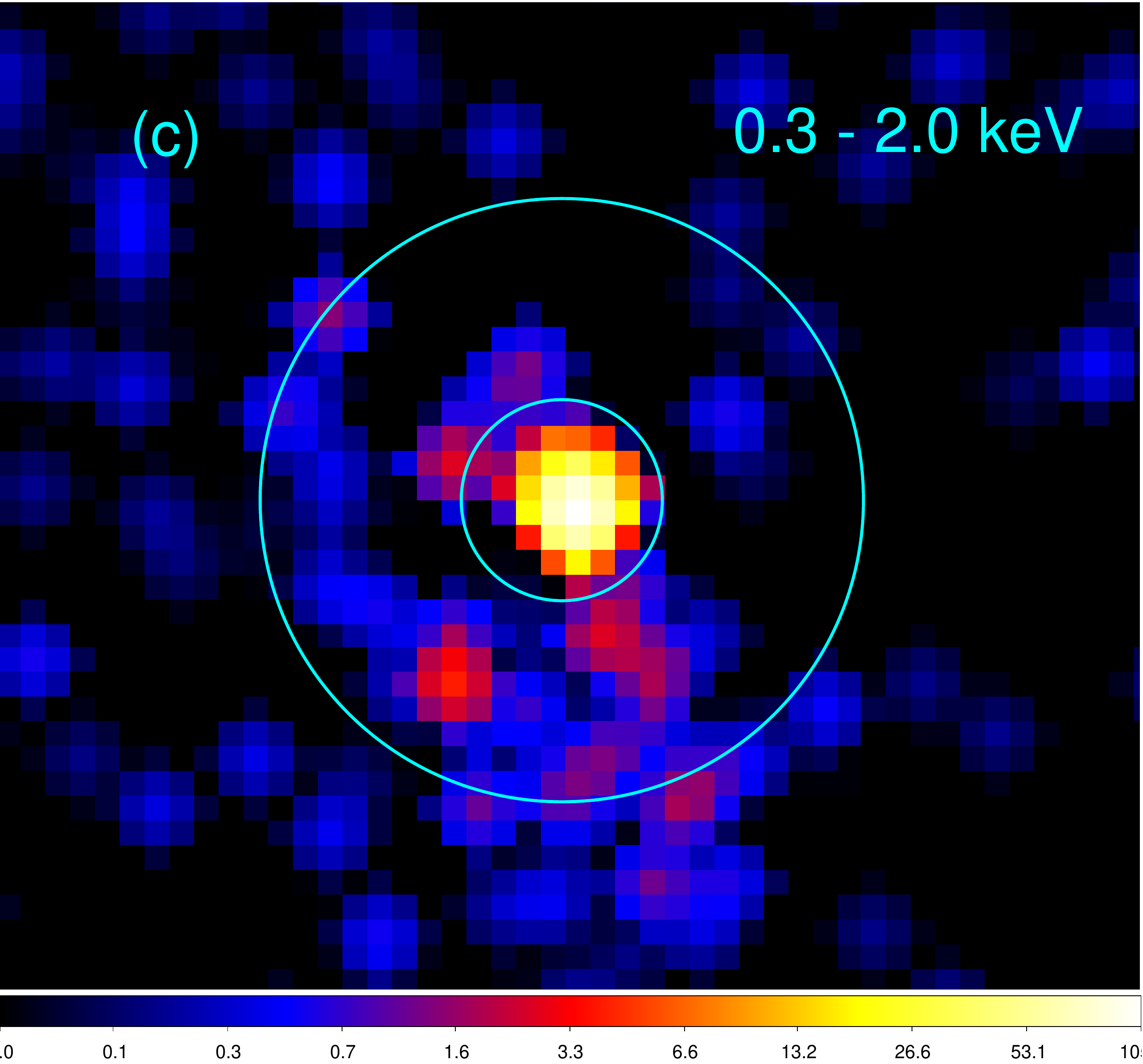}{0.3\textwidth}{}
	}
	\gridline{\fig{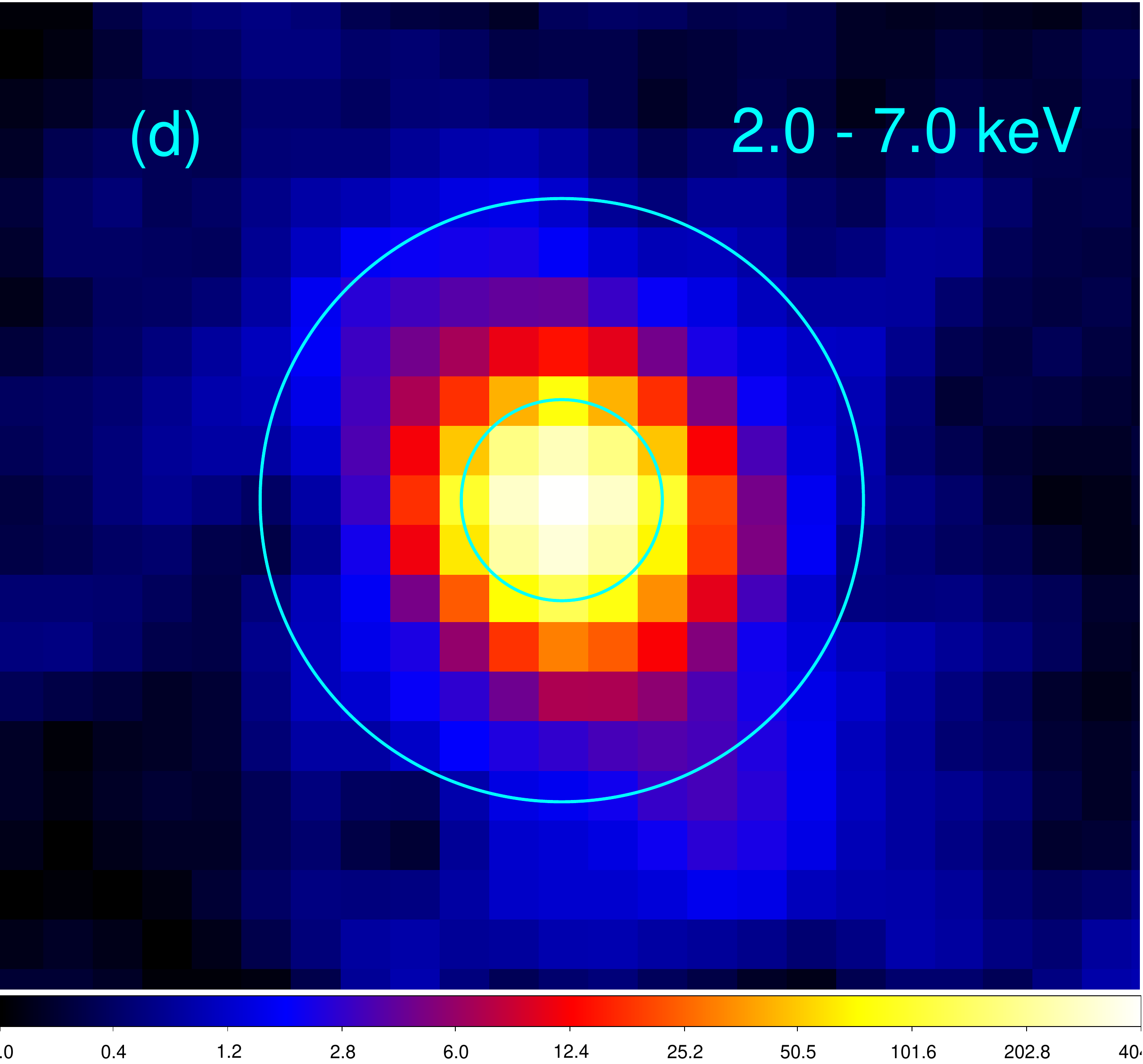}{0.3\textwidth}{}
		\fig{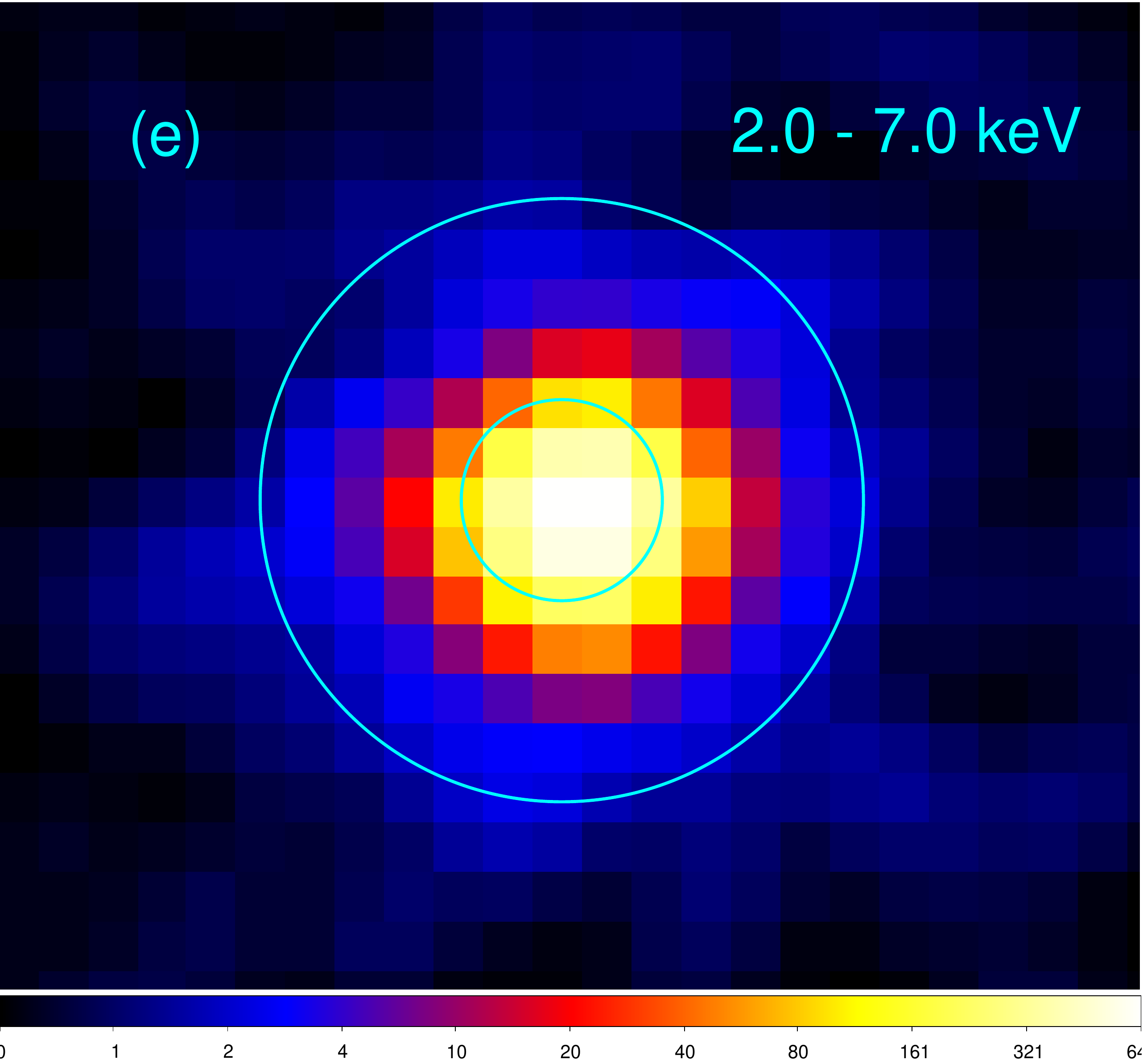}{0.3\textwidth}{}
		\fig{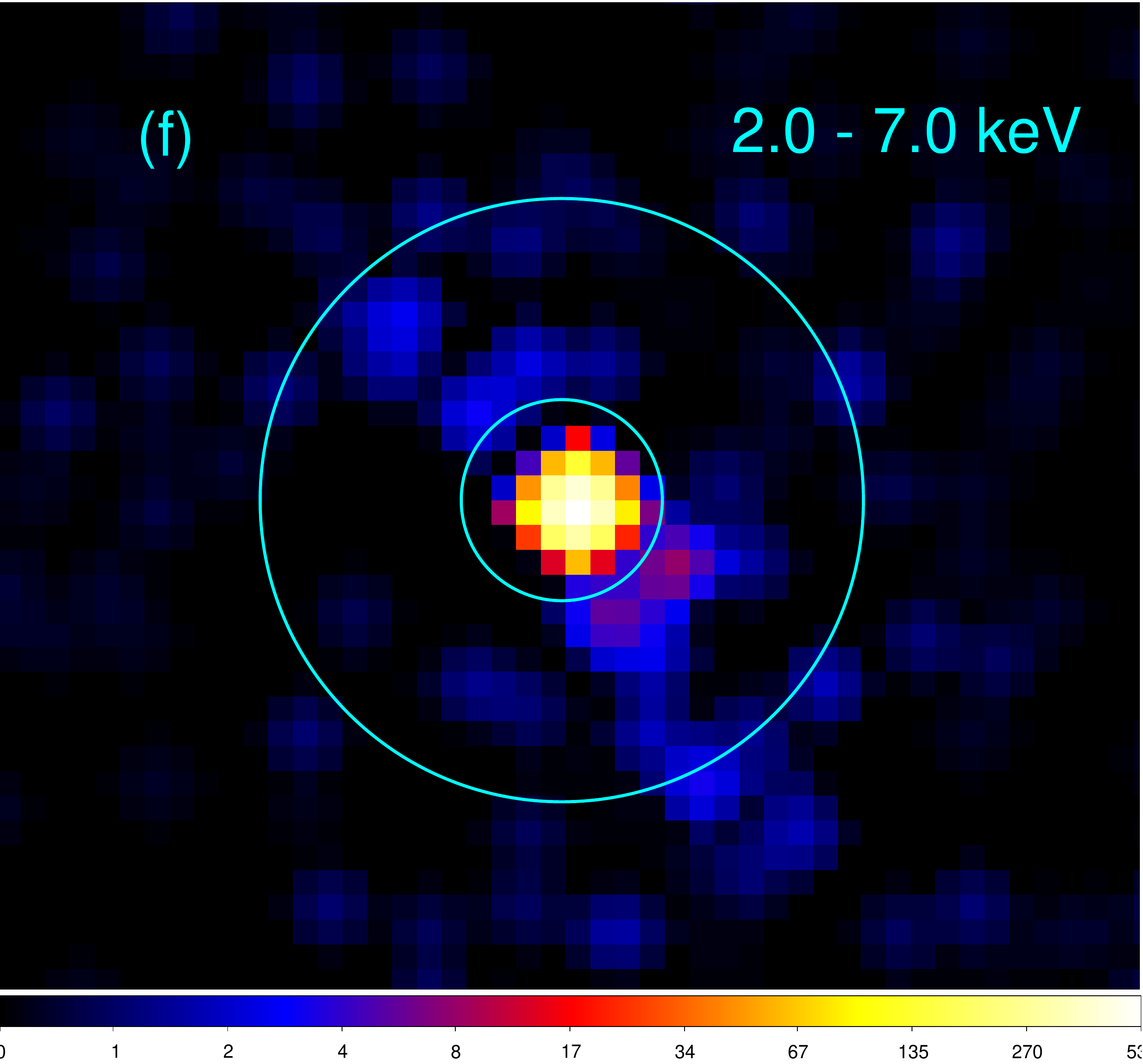}{0.3\textwidth}{}
	}
	\caption{(a) Combined HETG zeroth-order image of 0.3--2.0 keV band. 
	(b) PSF image of 0.3--2.0 keV band. 
	(c) Deconvolved image of 0.3--2.0 keV with binning factor is 0.5. 
	(d) Combined HETG zeroth-order image of 2.0--7.0 keV band. 
	(e) PSF image of 2.0--7.0 keV band. 
	(f) Deconvolved image of 2.0--7.0 keV with binning factor is 0.5. 
	All images are logarithmic scaled and smoothed with a Gaussian kernel of $\sigma = 1.0$ pixels. 
	The inner and outer radii of the cyan annulus in all panels are 1\arcsec\ and 3\arcsec\ respectively.
	The top is the north and the left is the east.
	\label{fig-0th-image}}
\end{figure*}

In our combined HETG zeroth-order images, there exists asymmetric extended X-ray emission in the near-nuclear region (Figure \ref{fig-0th-image}a \& d), especially in the soft band images (0.3--2 keV). 
To investigate the extended emission, we simulated the Point Spread Function (PSF) file with {\em Chandra Ray Tracer} (ChaRT) \footnote{\url{http://cxc.cfa.harvard.edu/ciao/PSFs/chart2/index.html/}}, using the spectrum of the X-ray core ($\leqslant$ 1\arcsec, considering the pile-up effect) as an input model.
The X-ray core's spectrum was fitted by a model that is \textit{pileup$\times$(phabs$\times$powerlaw + phabs$\times$(zgaussian + powerlaw)}, and the best-fit parameters can be found in Table~\ref{table-1arcsec}.
The photon index of \textit{powerlaw} is fixed at 1.86 according to \cite{2005ApJ...628..113C}, which is also consistent with a typical value (1.7--2.1) for Seyfert galaxy\citep[][]{2009A&A...495..421B,2009MNRAS.399.1597S}.
The output files obtained from ChaRT were converted to event files with the MARX {\em Chandra} simulator\footnote{\url{https://space.mit.edu/cxc/marx/}}. 
The resulting PSF images are shown in Figures~\ref{fig-0th-image}b and~\ref{fig-0th-image}e.   
We then used the {\tt arestore} script\footnote{\url{https://cxc.cfa.harvard.edu/ciao4.10/ahelp/arestore.html}} to enhance the image resolution with deconvolution techniques. 
The deconvolved images are shown in Figures~\ref{fig-0th-image}c and~\ref{fig-0th-image}f. 
Soft emissions extend to the south and southeast in Figure~\ref{fig-0th-image}c, whereas the hard emission is nearly along the direction of major axis of galaxy (Figure~\ref{fig-0th-image}f). 

\begin{figure}[htbp!]
	\includegraphics[width=1.1\columnwidth]{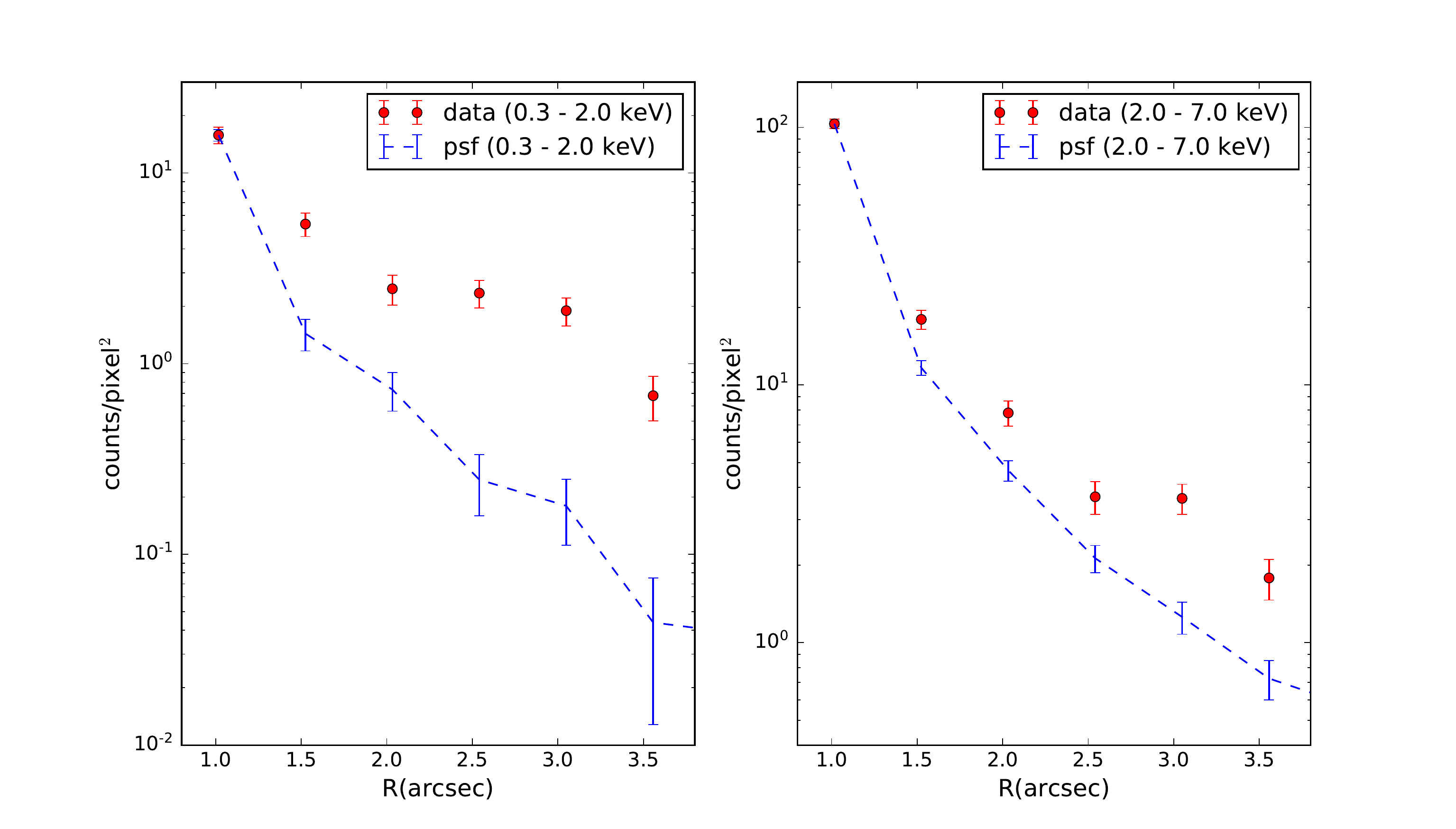}
	\caption{Radial profile images of 0.3--2.0 keV band and 2.0--7.0 keV band respectively. 
	Innermost points of the simulated PSF (dashed line) are normalized to the observation data (red points) in two panels.\label{fig-rprofile}}
\end{figure}

\begin{figure}[htbp!]
	\includegraphics[width=\columnwidth]{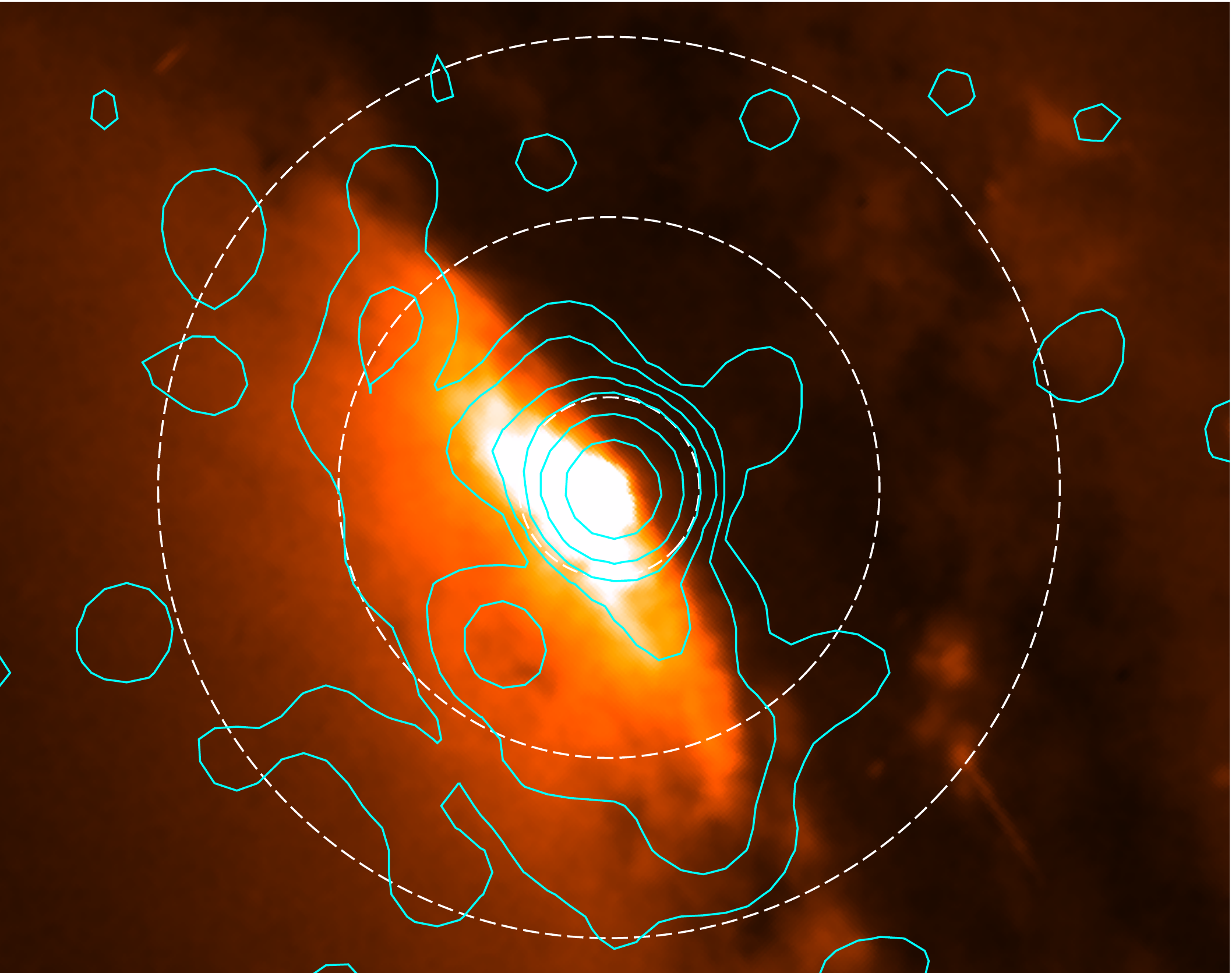}
	\caption{The HST/WFPC2 $V$-band image using the F606W filter \citep{1998ApJS..117...25M}. Overlaid are the X-ray contours of the deconvolved image in $\rm 0.3$--$2.0\ keV$ band ($\sigma = 4$, the lowest level is 0.09 $\rm counts$ $\rm pixel^{-1}$ and contour values increase by a factor of 2.9) from Figure~\ref{fig-0th-image}c.
	Three white dashed circles are centered on the position of the X-ray core and have radii of 1\arcsec, 3\arcsec, and 5\arcsec, respectively. 
	\label{fig-vband-sx}}
\end{figure}

We further compare the radial surface brightness profiles for the expected PSF and the observed data in Figure \ref{fig-rprofile}.   
The innermost data point ($\sim$1\arcsec) of the PSF is normalized to the corresponding innermost observed data.  There exists an apparent excess in the soft band $\sim$1\arcsec--3.5\arcsec\ from the nucleus. 
In the hard band, the data is also higher than the PSF $\sim$1\arcsec--3.5\arcsec\ from the nucleus, but not as apparent as in the soft band. 
In Figure \ref{fig-vband-sx}, the {\em Hubble Space Telescope} (HST) $V$-band image \citep[][]{1998ApJS..117...25M} superposed with the contours of soft X-ray emission (0.3--2.0 keV) is shown.
The dust lane in the northwest of the $V$ band image taken by HST is consistent with the absence of soft X-ray emission due to significant absorption in the same region. 

\begin{table}
	\centering
	\begin{tabular}{ccc}
		\hline
		\hline
		Component&Parameter&Best-fit value\\
		\hline
		
		pileup&frame time (s)&3.2 (f)\\
		
		&max\_ph&5 (f) \\
		
		&g0&1 (f)\\
		
		&alpha&1 (f)\\
		
		&PSF fraction&0.85 (f)\\
		
		phabs$_{1}$&$N_{\rm H,1}\ (\times 10^{22}\ cm^{-2})$&0.66$^{+0.05}_{-0.05}$\\
		
		$\rm powerlaw_{1}$&$\Gamma_{1}$&1.86 (f)\\
		
		&$\rm Norm\ (\times 10^{-4})$&6.77$^{+0.51}_{-0.53}$\\
		
		phabs$_{2}$&$N_{\rm H,2}\ (\times 10^{22}\ cm^{-2})$&8.53$^{+2.77}_{-1.84}$\\
		
		$\rm zGauss$&$E$ (keV)&6.38$^{+0.01}_{-0.01}$\\
		
		&$\rm Norm\ (\times 10^{-5})$&3.66$^{+0.34}_{-0.34}$\\
		
		$\rm powerlaw_{2}$&$\Gamma_{2}$&1.86 (f)\\
		
		&$\rm Norm\ (\times 10^{-4})$&7.66$^{+0.73}_{-0.65}$\\
		
		\hline
		-&$\chi^{2}/d.o.f$&141.1/139\\
		\hline	
	\end{tabular}
	\caption{
	Parameters of the model used to fit the nuclear spectrum ($\leqslant$ 1\arcsec). 
	$N_{\rm H}$ is the column density of the hydrogen, and the subscript 1 or 2 correspond to the first or second \textit{phabs} component. 
	$\Gamma_{1}$ and $\Gamma_{2}$ are the photon index of the two \textit{powerlaw} component, respectively. 
	The center energy of the primary Fe K$\alpha$ emission line is shown by $E$. 
	Norm is the normalization parameter.
	Fixed parameters are indicated by ``(f)''.
	}
	\label{table-1arcsec}
\end{table}

\subsection{Spectroscopy} \label{subsec:Spectroscopy}

\begin{figure}[htbp!]
	\includegraphics[width=0.7\columnwidth,angle=270]{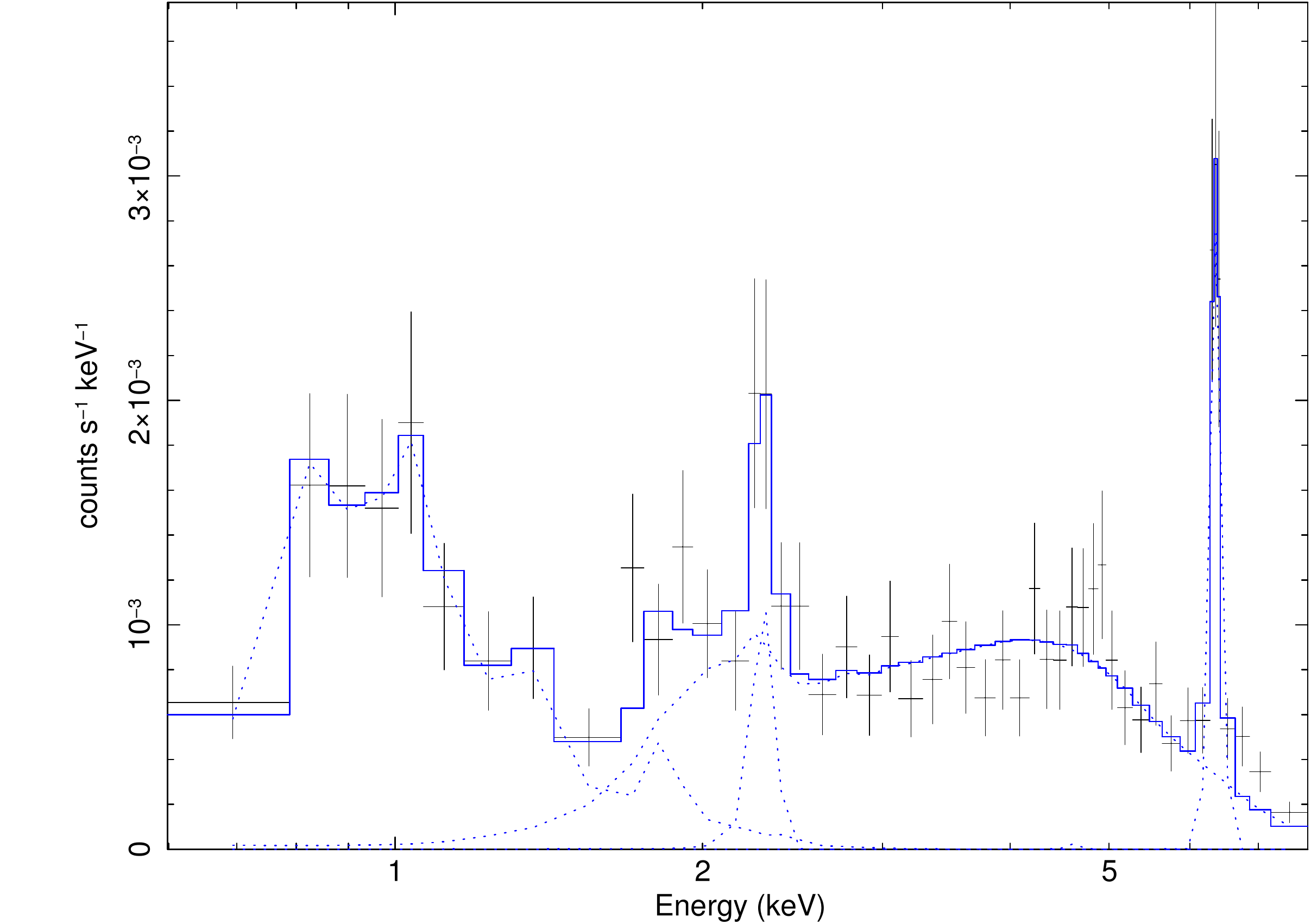}
	\caption{{\em Chandra} ACIS spectrum (in the observed frame) extracted from the three available observations. 
	The extraction region is shown as the cyan annulus in Figure \ref{fig-0th-image}. 
	The black points, blue line, and blue dotted line denote the data, synthetic model, and model components, respectively.
	Parameters of the best fit collisionally-ionized model, i.e. ${phabs\times(\textit{apec}) + phabs\times(powerlaw + zGauss + zGauss)}$, are shown in Table \ref{table-apec}.
	\label{fig-spec-apec}}
\end{figure}

The extended soft X-ray emission of NGC 2992 in the 1\arcsec--3\arcsec\ region could be due to winds driven by AGN or starburst, or AGN-photoionization. 
To investigate the origin of this soft X-ray emission, we extracted a spectrum from the annulus shown in Figure \ref{fig-0th-image}. 
In Figure \ref{fig-spec-apec}, the spectrum fitted with three components (optically thin thermal plasma, power-law continuum and two emission lines) from 0.5 to 8 keV is shown. 
Details of the model parameters are listed in Table \ref{table-apec}. 

In Figure \ref{fig-rprofile}, the hard band extended emission is less significant (described later in the Section \ref{sec:hard}). 
An absorbed power-law component with a photon index $\Gamma$ fixed at 1.86 was used to fit the hard X-ray continuum of the extracted spectrum, assuming that the hard band X-ray emission is dominated by the PSF from the nucleus ($\leqslant$ 1\arcsec) of NGC 2992.
An absorbed thermal plasma model \textit{apec} \citep{2001ApJ...556L..91S} is added to describe the soft emission, with abundance fixed at solar values $\rm Z_{\astrosun}$.  
In addition, two emission lines are obvious in the spectrum with centroid energy at 2.31 keV and 6.4 keV, and are accounted for with two gaussian components. 
The first line corresponds to the $\rm S\ K\alpha$ that had previously been detected in the first-order, co-added HEG and MEG spectrum of the same data by \cite{2017ApJ...840..120M}, and the other is the $\rm Fe\ K\alpha$, which is the fluorescent neutral iron line. 
The line widths (velocity dispersion) of these two emission lines were fixed at 0.01 keV since they were both unresolved narrow lines in the spectrum. 
The residual at $\sim4.9\rm\,keV$ indicates possible presence of another emission line, which is further analyzed later in the Section \ref{sec:hard}.

The best fit absorption column and temperature of the hot plasma of the \textit{apec} component is $N_{\rm H,1} = 0.84 \times10^{22}\rm\ cm^{-2}$ and $kT = 0.33$ keV, respectively (Table \ref{table-apec}). 
\cite{2005ApJ...628..113C} was able to extract spectrum from the circum-nuclear region ($\sim$3\arcsec--18\arcsec) and the resulting best fit temperature of thermal plasma was 0.5 keV with an unphysically low abundance ($Z <0.03\rm Z_{\astrosun}$). However, if they fix the abundance at $Z=\rm Z_{\astrosun}$, the temperature would change to $kT = 0.31$ keV, which is consistent with our result. 
We also attempted to allow the abundance to vary, whereas the fitting results show no improvement.  The observed luminosity of the \textit{apec} component is $\sim 7.0\times10^{39}\rm\ erg\ s^{-1}$ which is lower than the luminosity for soft excess $\sim 10^{40}\rm\ erg\ s^{-1}$ predicted in \cite{2005ApJ...628..113C}. 

\begin{figure}[htbp!]
    \includegraphics[width=0.7\columnwidth,angle=270]{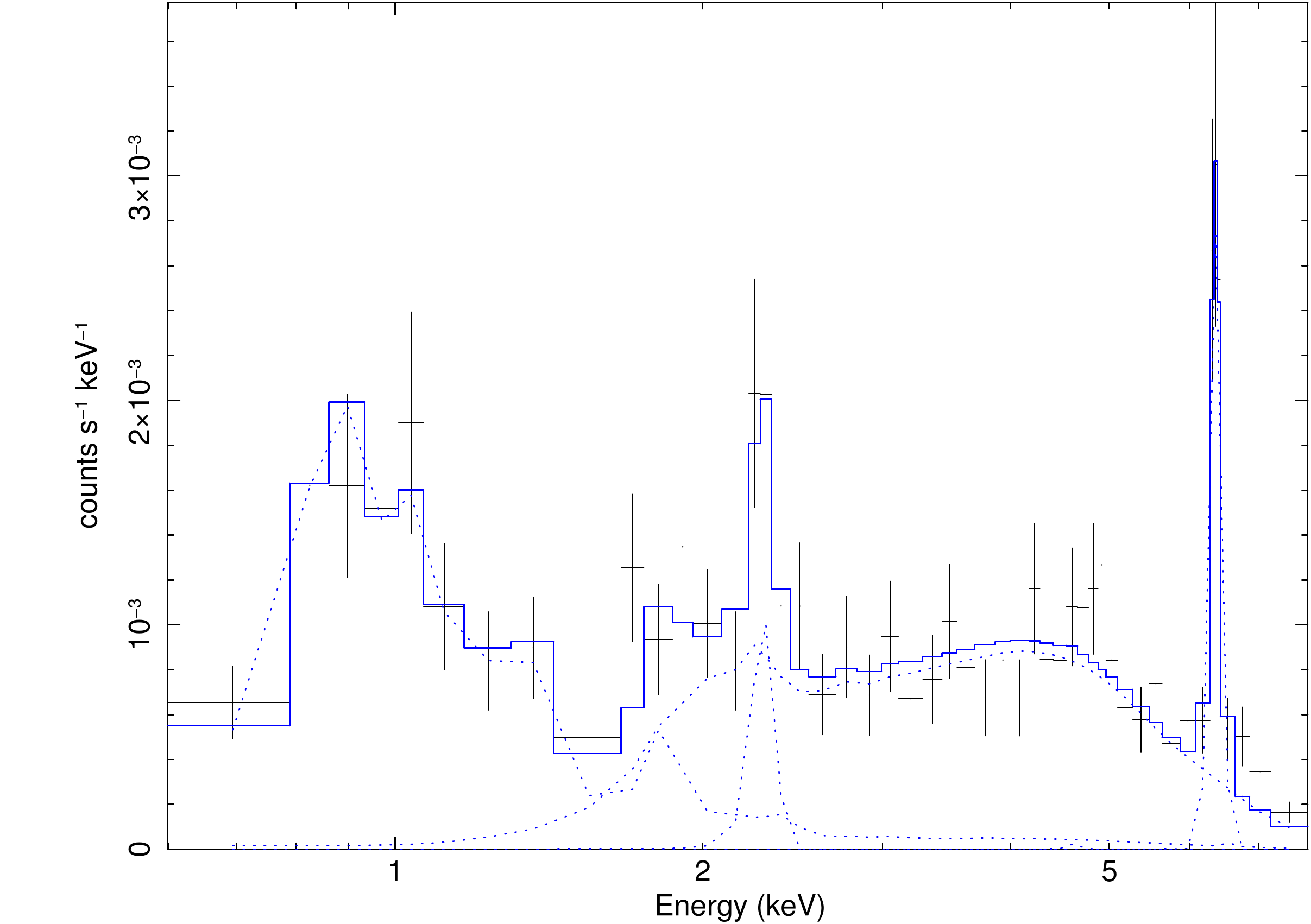}
	\caption{The same spectrum as Figure \ref{fig-spec-apec} whereas the \textit{apec} component is replaced by a photoionized model which generated by CLOUDY.
	Parameters of the best fit model are shown in Tabel \ref{table-cloudy}.
	\label{fig-spec-cloudy}}
\end{figure}

To explore the possibility that the soft excess could be originated from photoionization, we used CLOUDY \citep[version 17.01;][]{2017RMxAA..53..385F} to model a photoionized component replacing the \textit{apec} component. The parameters for generating the model grid are $\rm log U=[-2.00:2.50]$ and $\rm log N_{H} (cm^{-2})=[16.00:22.00]$ both with steps of 0.1 dex. Only the reflected spectrum generated from the illuminated face of the cloud has been taken into consideration in our model. Figure \ref{fig-spec-cloudy} shows the data with model components, and the parameters of the best-fit model are presented in table \ref{table-cloudy}. 
The best-fit parameter of the ionization is $\rm log U=0.91$, nevertheless the associated column density cannot be well constrained with only an upper limit $\rm log N_{H} (cm^{-2}) \leqslant 20.0$. The observed luminosity of CLOUDY model $\sim 7.7\times10^{39}\rm\ erg\ s^{-1}$ is comparable to the \textit{apec} component described above.

\begin{table}
	\centering
	\begin{tabular}{ccc}
		\hline
		\hline
		Component&Parameter&Best-fit value\\
		\hline
		
		$\rm phabs_{1}$&$N_{\rm H,1}\ (\times 10^{22}\ cm^{-2})$&0.84$^{+0.16}_{-0.16}$\\
		
		apec&$kT$ (keV)&0.33$^{+0.21}_{-0.08}$ \\
		
		&$\rm Norm\ (\times 10^{-4})$&4.66$^{+8.06}_{-3.07}$\\
		
		$\rm phabs_{2}$&$N_{\rm H,2}\ (\times 10^{22}\ cm^{-2})$&3.27$^{+1.02}_{-0.53}$\\
		
		powerlaw&$\Gamma$&1.86 (f)\\
		
		&$\rm Norm\ (\times 10^{-4})$&1.14$^{+0.10}_{-0.08}$\\
		
		$\rm zGauss_{1}$&$E_{1}$ (keV)&2.31$^{+0.02}_{-0.02}$\\
		
		&$\rm Norm\ (\times 10^{-6})$&3.98$^{+2.79}_{-1.74}$\\
		
		$\rm zGauss_{2}$&$E_{2}$ (keV)&6.40$^{+0.01}_{-0.01}$\\
		
		&$\rm Norm\ (\times 10^{-6})$&5.06$^{+0.75}_{-0.75}$\\
		\hline
		-&$\chi^{2}/d.o.f$&32.6/41\\
		\hline	
	\end{tabular}
	\caption{Parameters of the model used in figure \ref{fig-spec-apec}. 
	$N_{\rm H}$ is the column density of the hydrogen, and the subscript 1 or 2 correspond to the first or second \textit{phabs} component. 
	$kT$ indicates the temperature of thermal plasma and $\Gamma$ denotes the photon index of the \textit{powerlaw}. 
	Centroid energies of two emission lines are shown by $E_{1}$ and $E_{2}$ respectively. 
	}
	\label{table-apec}
\end{table}

\begin{table}
	\centering
	\begin{tabular}{ccc}
		\hline
		\hline
		Component&Parameter&Best-fit value\\
	\hline
	
	phabs$_{1}$&$N_{\rm H,1}\ (\times 10^{22}\rm\, cm^{-2})$&0.40$^{+0.10}_{-0.17}$\\
	
	CLOUDY&${\rm log} U$&0.83$^{+0.23}_{-0.03}$ \\
	
	&${\rm log} N_{\rm H} (\rm cm^{-2})$&$\leqslant 20.0$ \\
	
	phabs$_{1}$&$N_{\rm H,2}\ (\times 10^{22}\rm\, cm^{-2})$&3.17$^{+0.92}_{-0.67}$\\
	
	powerlaw&$\Gamma$&1.86 (f)\\
	
	&$\rm Norm\ (\times 10^{-4})$&1.08$^{+0.10}_{-0.08}$\\
	
	zGauss$_{1}$&$E_{1}$ (keV)&2.31$^{+0.02}_{-0.02}$\\
	
	&$\rm Norm\ (\times 10^{-6})$&3.83$^{+2.34}_{-1.76}$\\
	
	zGauss$_{2}$&$E_{2}$ (keV)&6.40$^{+0.01}_{-0.01}$\\
	
	&$\rm Norm\rm (\times 10^{-6})$&5.05$^{+0.75}_{-0.75}$\\
	\hline
	-&$\chi^{2}/d.o.f$&34.3/40\\
	
	\hline	
	\end{tabular}
	\caption{Parameters of the model used in Figure \ref{fig-spec-cloudy}. 
	${\rm log} U$ is the ionization parameter of CLOUDY model and its column density of hydrogen is indicated by ${\rm log} N_{\rm H}$. 
	Other parameters are the same as Table \ref{table-apec}.}
	\label{table-cloudy}
\end{table}

\section{Two Emission Lines in the Hard Band\label{sec:hard}}

\subsection{Spatial and spectral analysis} \label{subsec:SSA}

\begin{figure*}[ht!]
	\gridline{\fig{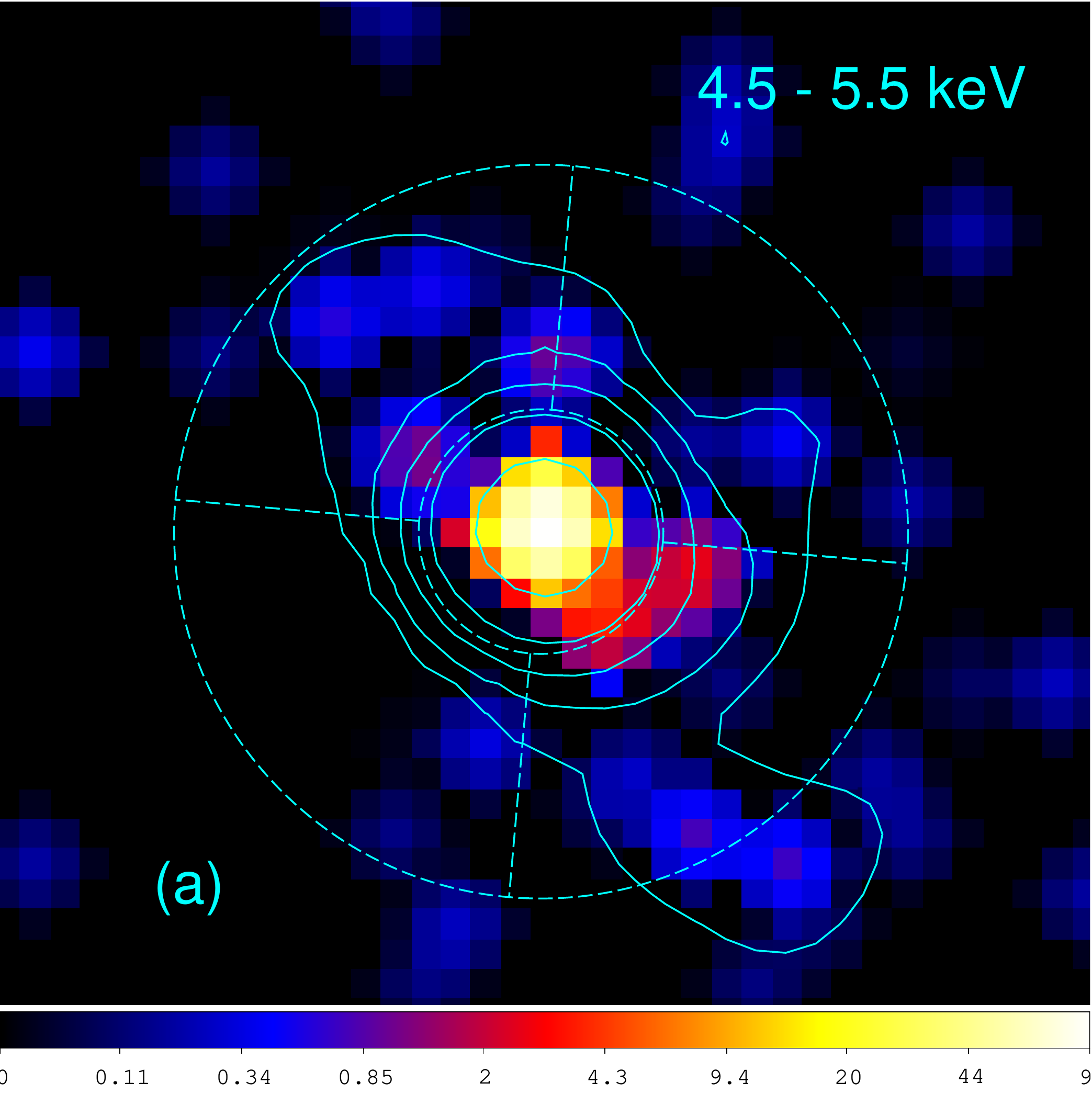}{0.3\textwidth}{}
		\fig{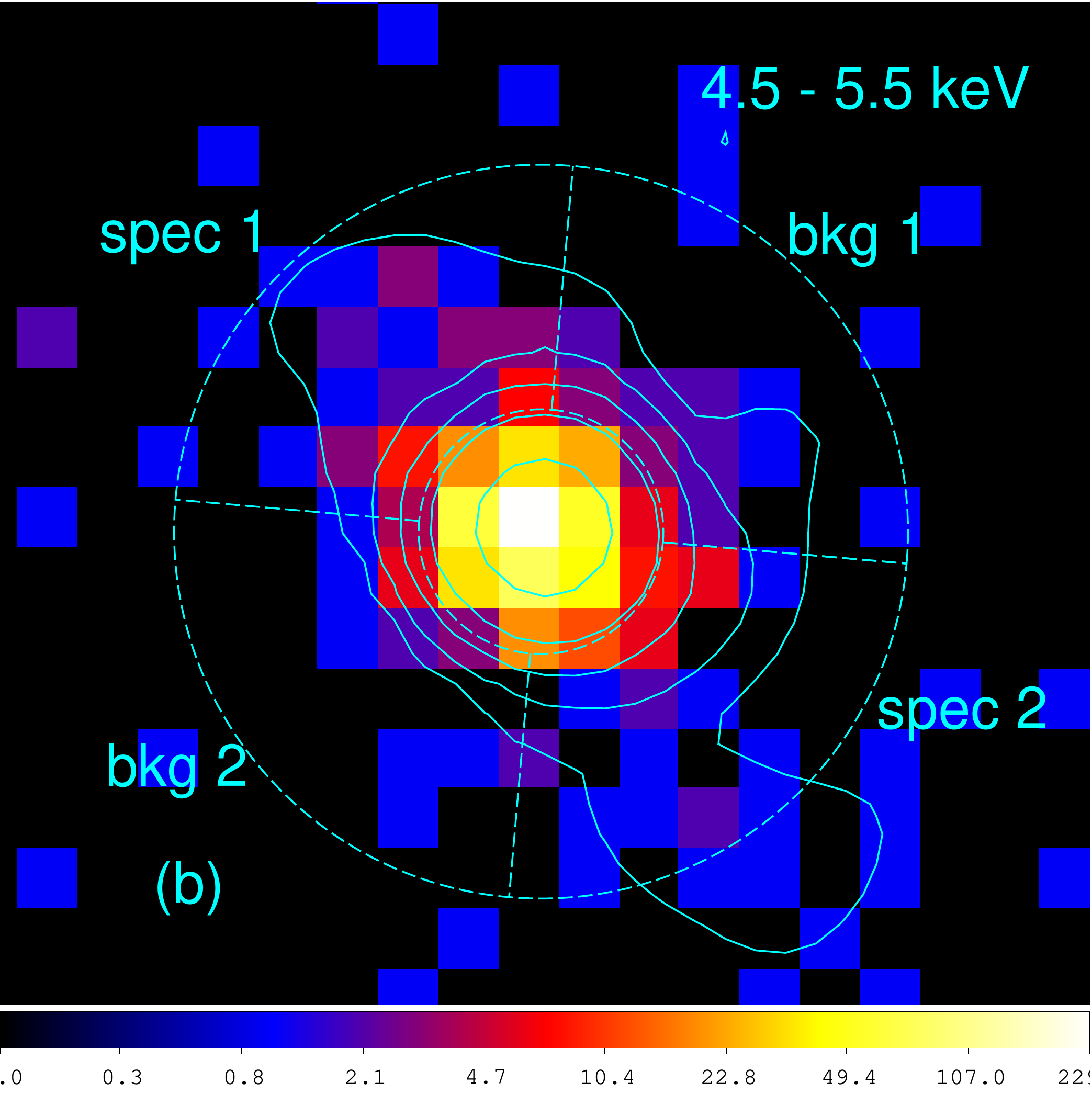}{0.3\textwidth}{}
		\fig{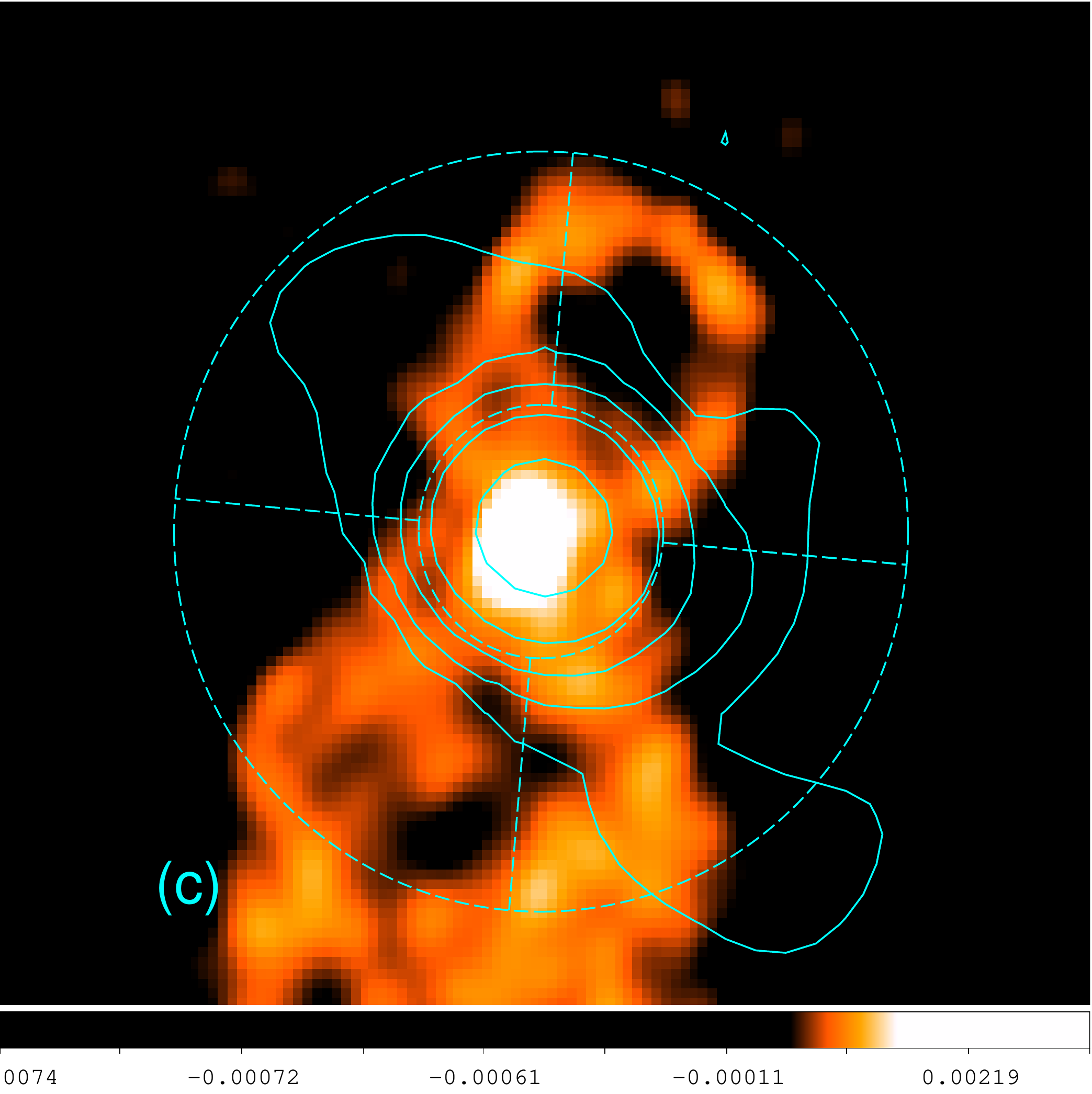}{0.3\textwidth}{}
	}
	\gridline{\fig{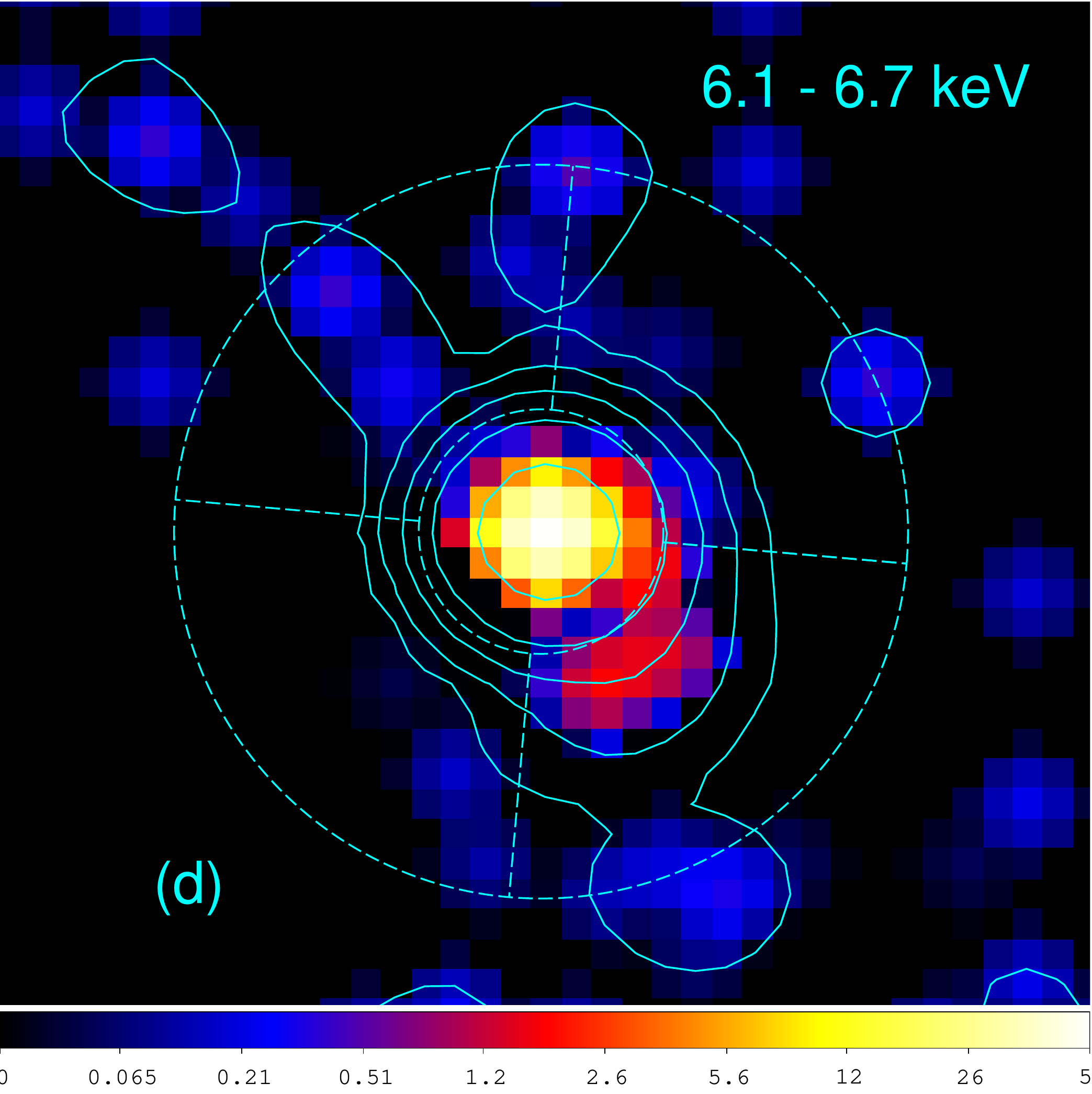}{0.3\textwidth}{}
		\fig{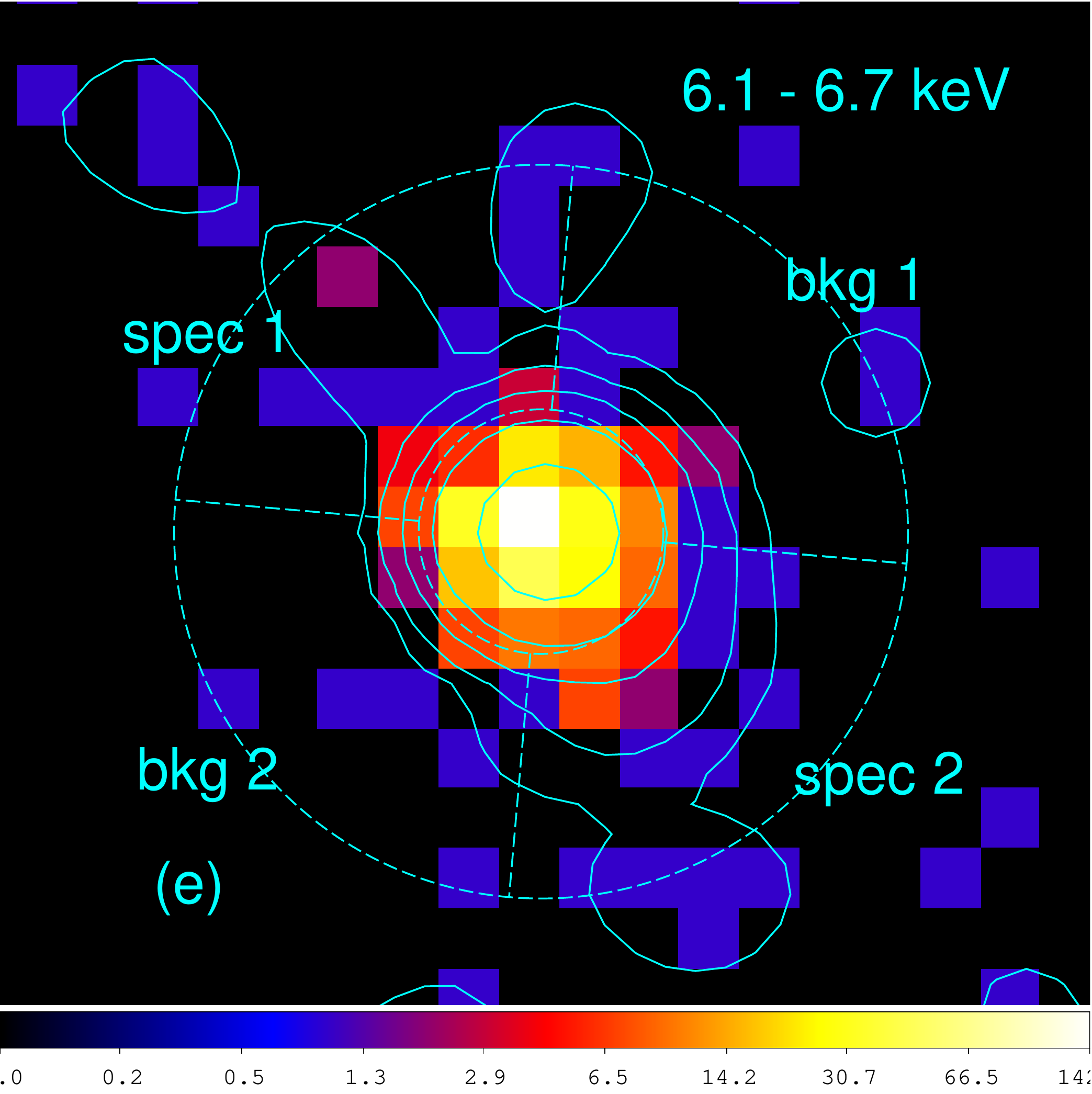}{0.3\textwidth}{}
		\fig{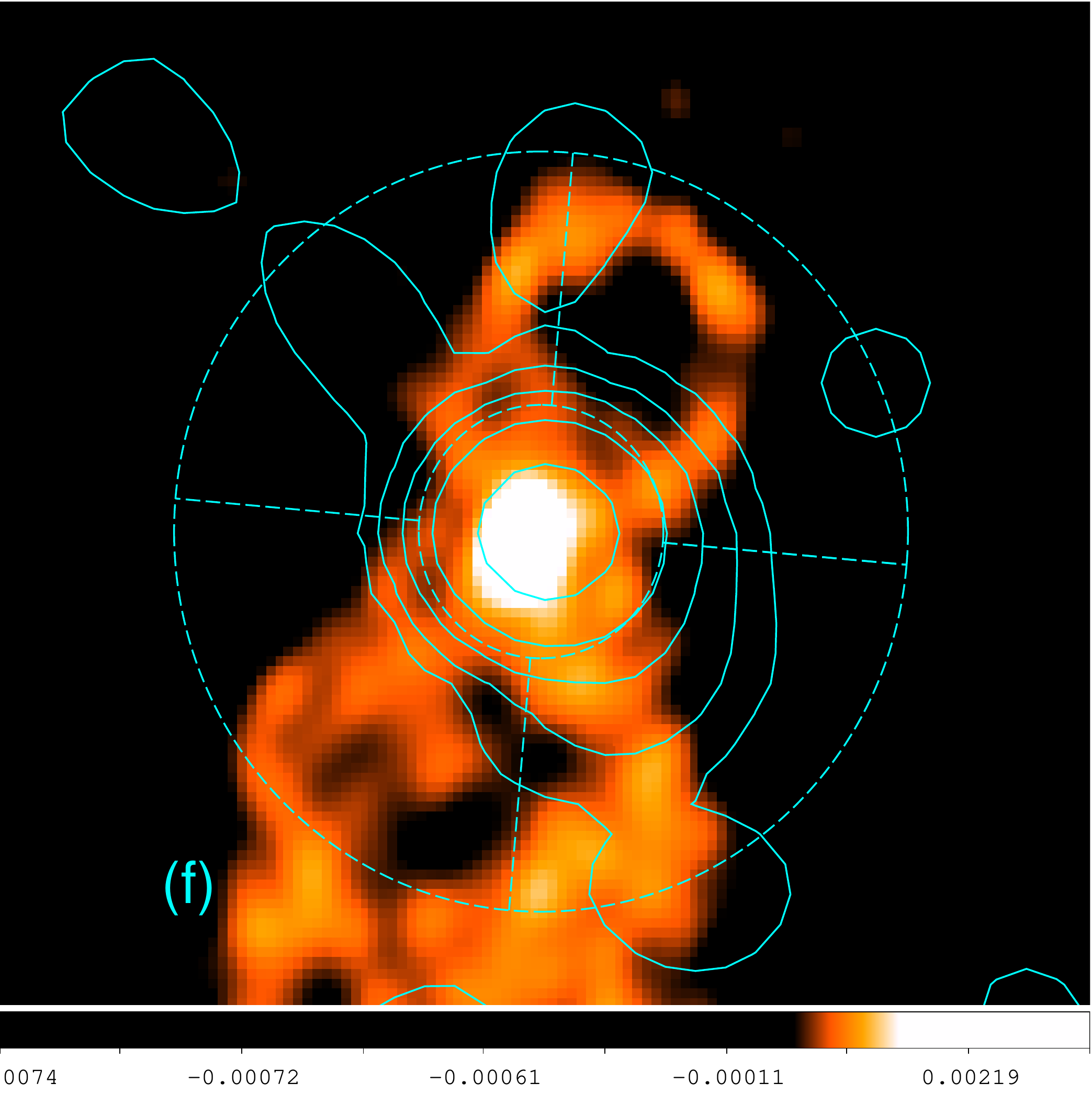}{0.3\textwidth}{}
	}
	\caption{(a) Deconvolved image (smoothed with a Gaussian kernel and $\sigma = 1.5$ pixels) and contours ($\sigma = 4$, lowest level is 0.1 $\rm counts\ pixel^{-1}$) of 4.5--5.5 keV band, with the binning factor of 0.5. 
	(b) Combined HETG zeroth-order image of 4.5--5.5 keV band with contours of (a). 
	(c) VLA 6 cm radio image (from NED) with contours of (a). 
	(d) Deconvolved image (smoothed with $\sigma = 1.5$ pixels) and contours ($\sigma = 4$, lowest level is 0.1 $\rm counts\ pixel^{-1}$) of 6.1--6.7 keV band, with binning factor is 0.5. 
	(e) Combined HETG zeroth-order image of 6.1--6.7 keV band with contours of (d). 
	(f) VLA 6 cm radio image with contours of (d). 
	All images and contours are logarithmic scaled. 
	The inner radius and the outer radius of dashed-line quadrant regions in all panels is 1\arcsec\ and 3\arcsec, respectively. 
	\label{fig-hard-image}}
\end{figure*}

\begin{figure}[htbp!]
	\includegraphics[width=1.2\columnwidth]{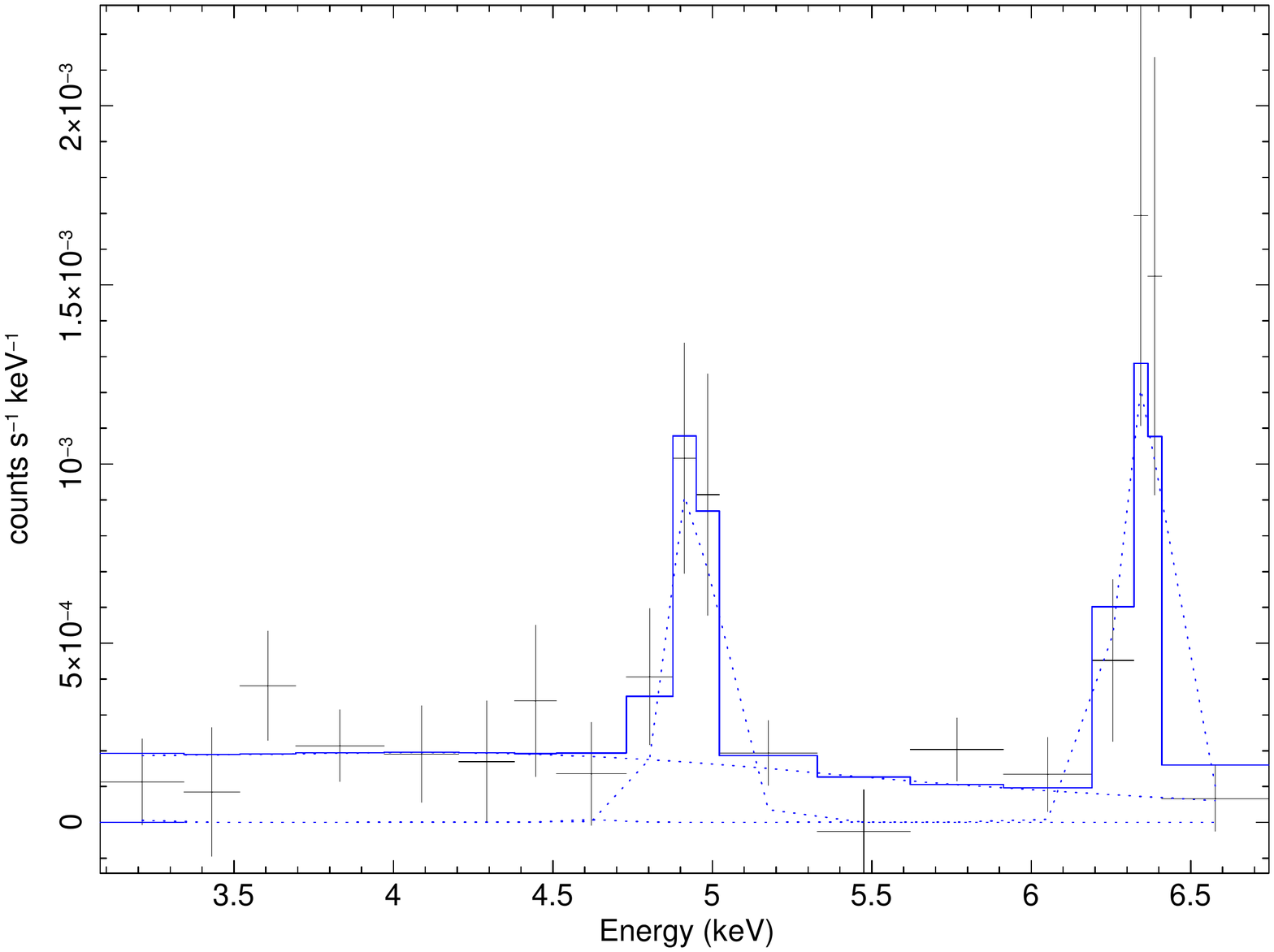}
	\caption{Spectrum extracted from spec 2 region in Figure \ref{fig-hard-image}.
	The black points, blue line, and blue dotted line denote the data, synthetic model, and model components (a \textit{powerlaw} and two \textit{zGauss}), respectively.
	\label{fig-spec-2lines}}
\end{figure}

There also exists extended hard X-ray emission along with the disk of NGC 2992 shown in Figure \ref{fig-0th-image}f which can be corroborated by the excess of hard counts in the right panel of Figure \ref{fig-rprofile}.
In order to investigate the hard X-ray extended emission (2--7 keV), we divided the annulus (1\arcsec--3\arcsec) into four regions (Figure \ref{fig-hard-image}) which were labeled as ``bkg 1", ``spec 1", ``bkg 2", and ``spec 2", given that the hard emission appears to extend along the direction of the major axis of NGC 2992 galaxy. 
To reduce the influence of nuclear PSF, we used bkg 1 and bkg 2 sectors as background regions, assuming emissions from these two sectors were mainly originated from the PSF of the nucleus, whereas spec 1 and spec 2 regions were used for spectral extraction. 
Spectra from spec 1 and spec 2 regions were extracted, respectively. 
The continuum of the spectra can be reduced to a low level (observed flux $F_{3-7\rm\,keV}\sim3.4\times10^{-14}\rm\,erg\,cm^{-2}\,s^{-1}$) using the bkg 1 and bkg 2 regions as the background, hence the hard emission from extended structure can be detected more distinctly, e.g. the normalization of the 4.97 keV line increase from $9.3^{+0.34}_{-0.31}\times10^{-7}$ (using the circular region in Sect.~\ref{sec:data} as background) to $11.4^{+0.33}_{-0.32}\times10^{-6}$.
Figure~\ref{fig-spec-2lines} shows the spectrum extracted from the spec 2 region and the model fit (an absorbed \textit{powerlaw} model with two emission lines) and the best-fit parameters are shown in Table \ref{table 3}. 
Two apparent emission lines centered at energy of 4.97 keV and 6.39 keV were detected in the spectrum extracted from the spec 2 region (Figure~\ref{fig-spec-2lines}).
The 6.39 keV line is likely to be the $\rm Fe\ K\alpha$, while the 4.97 keV line cannot be identified to a commonly observed X-ray emission line.  
The line widths of these two emission lines was fixed at 0.01 keV, since they could not be resolved at the spectral resolution of the CCD. 
In the spectrum of the spec 1 region we cannot confirm the presence of these two lines due to the lower signal to noise ratio, but the upper limit normalization of the 4.97 keV emission line ($\sim1.3\times10^{-6}$) is consistent with this line in the spec 2 region ($1.14\times10^{-6}$).
For the 6.39 keV emission line, the upper limit ($7.3\times10^{-7}$) is lower than this line in the spec 2 region ($2.25\times10^{-6}$).

\begin{figure}[htbp!]
	\includegraphics[width=1.1\columnwidth]{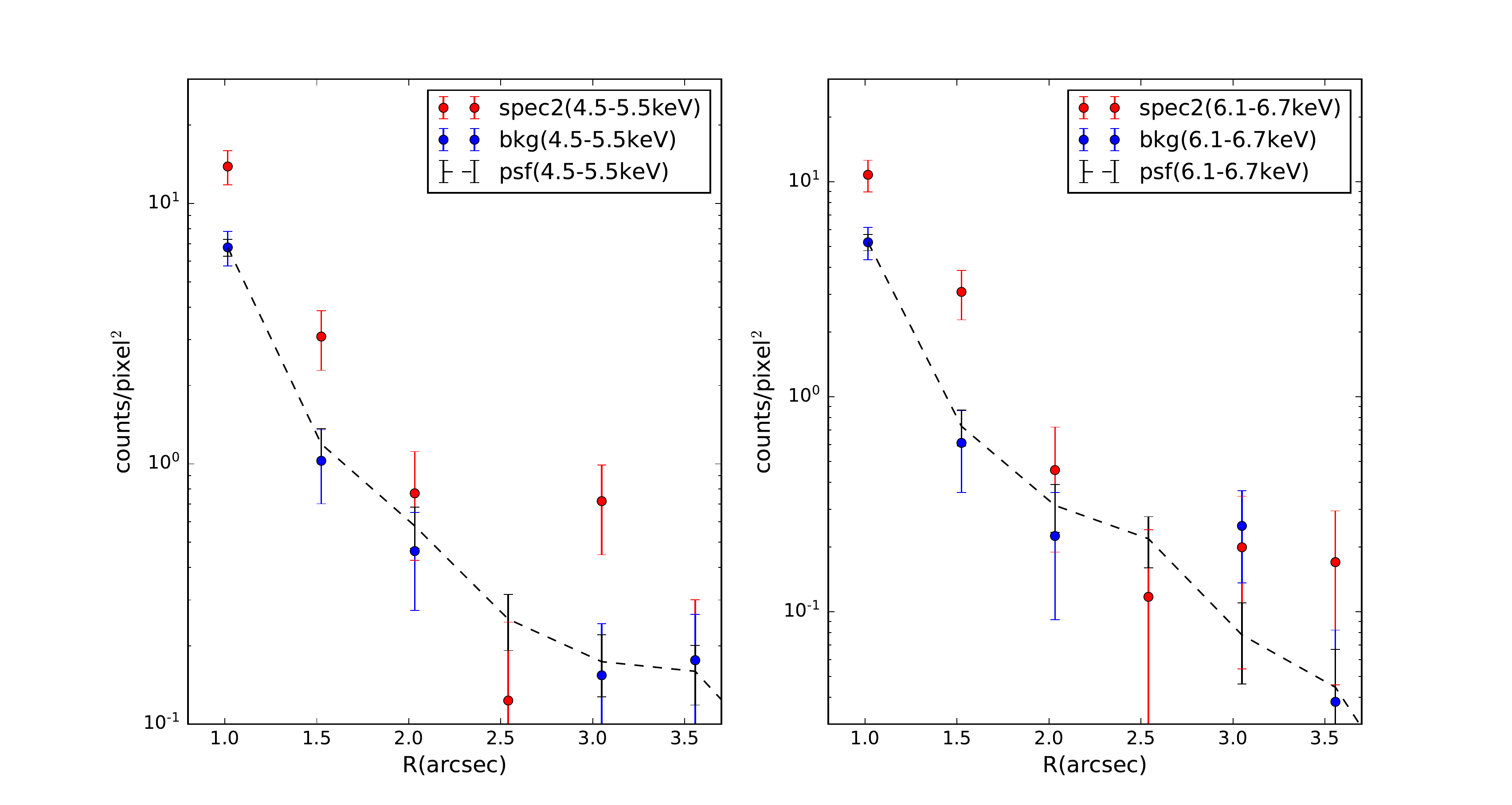}
	\caption{Images of radial profile. 
	The red circle, blue circle, and dashed lines denote the data from spec 2 region, bkg region, and PSF, respectively.
	Innermost points of simulated PSF are normalized to the innermost background (bkg) in two panels, respectively.\label{fig-rprofile-2lines}}
\end{figure}

The radial profiles of these regions in 4.5--5.5 keV and 6.1--6.7 keV band were plotted in Figure~\ref{fig-rprofile-2lines}, in which we normalized the innermost point of the PSF to the innermost background region. 
The consistent profiles of the PSF and the background reinforce our assumption about the origin of the hard continuum in these two energy bands. 
Majority of hard X-ray extended emission in the 4.5--5.5 and 6.1--6.7 keV bands is located in the spec 2 region $\leqslant2$\arcsec\ according to Figure~\ref{fig-hard-image}a \& d and Figure~\ref{fig-rprofile-2lines}. 

Comparing the 6 cm radio image from Very Large Array (VLA) to the X-ray contours (Figure \ref{fig-hard-image}c \& f), the $\rm Fe\ K\alpha$ emission appears to be located at the rims of the figure-eight loops, whereas the 4.97 keV line emission does not appear to align with the radio structure perfectly. Overall the morphology of these two line emission both extends along the major axis of the galaxy.

\begin{table}
	\centering
	\begin{tabular}{ccc}
		\hline 
		\hline
		Component&Parameter&Best-fit value/\\
				 & &Calculated result\\
		\hline
		
		phabs&$N_{\rm H}\ (\times 10^{22}\rm\, cm^{-2})$&3.2 (f)\\
		
		powerlaw&$\Gamma$&1.93$^{+1.16}_{-1.09}$\\
		
		&$\rm Norm\ (\times 10^{-5})$&2.70$^{+11.08}_{-2.19}$\\
		
		zGauss$_{1}$&$E_{1}$ (keV)&4.97$^{+0.03}_{-0.03}$\\
		
		&$\rm Norm\ (\times 10^{-6})$&1.14$^{+0.33}_{-0.32}$\\
		
		&$EW_{1}$ (keV)&0.92$^{+1.41}_{-0.17}$\\
		
		&$S_{1}$&99.9\%\\
		
		zGauss$_{2}$&$E_{2}$ (keV)&6.39$^{+0.02}_{-0.02}$\\
		
		&$\rm Norm\ (\times 10^{-6})$&2.25$^{+0.66}_{-0.61}$\\
		
		&$EW_{2}$ (keV)&2.93$^{+6.39}_{-0.50}$\\
		
		&$S_{2}$&99.9\%\\
		\hline
		-&$\chi^{2}/d.o.f$&8.7/13\\
		\hline		
	\end{tabular}
	\caption{Parameters of best-fit model, i.e. $phabs(powerlaw + zGauss + zGauss)$, in Figure \ref{fig-spec-2lines}. 
	$EW$ denotes the equivalent width of emission lines and $S$ is the statistical significance.}
	\label{table 3}
\end{table}

\subsection{Statistical significance of the emission lines} \label{subsec:SS}

As discussed in \cite{2002ApJ...571..545P}, standard likelihood ratio tests might lead to inaccurate results when calculating the statistical significance, thus likelihood ratio test was not adopted to evaluate the line detection. Instead, we produced Monte Carlo simulations of our data and retrieved statistical significance for two emission lines detected, which followed the method discussed in \cite{2010A&A...521A..57T} and \cite{2018MNRAS.478.5638M}. 10000 data sets of simulated spectra of the spec 2 region were produced, using the {\tt fakeit} command in XSPEC, with the same responses, background files, exposure times, and energy binning as the ones used for the observed data. The best-fit model presented in Table \ref{table 3} without emission lines was used to simulate a fake spectrum, and the best-fit parameters were recorded.  Next, this new model for the continuum was used to simulate the fake data again, which could reduce uncertainties on the null-hypothesis probability itself \citep{2004A&A...427..101P}. 
A new narrow Gaussian line was added to the model, and the normalization of this line was initially set to zero and free to vary. 
We set the centroid energy as a free parameter to vary from 3.0 to 7.0 keV in 50 eV steps and the resulting $\Delta\chi^{2}$ was recorded.  If $N$ was the number of datasets in which a lower $\Delta\chi^{2}$ than the threshold value was found and $S$ was the number of simulated spectra, we could calculate the estimated statistical significance of the detection from Monte Carlo simulations with $1-N/S$. 
Threshold values were set as $-$12.3 for the 4.97 keV emission line and $-$9.2 for the $\rm Fe\ K\alpha$ line, which were both derived from the fitting result of the real data.  The statistical significance of these two emission lines both are $\geqslant99.9\%$ (Table \ref{table 3}).

\section{Discussion} \label{sec:discussion}
\subsection{Extended Soft X-Ray Emission} \label{subsec:soft excess}

\begin{figure}[htbp!]
	\plotone{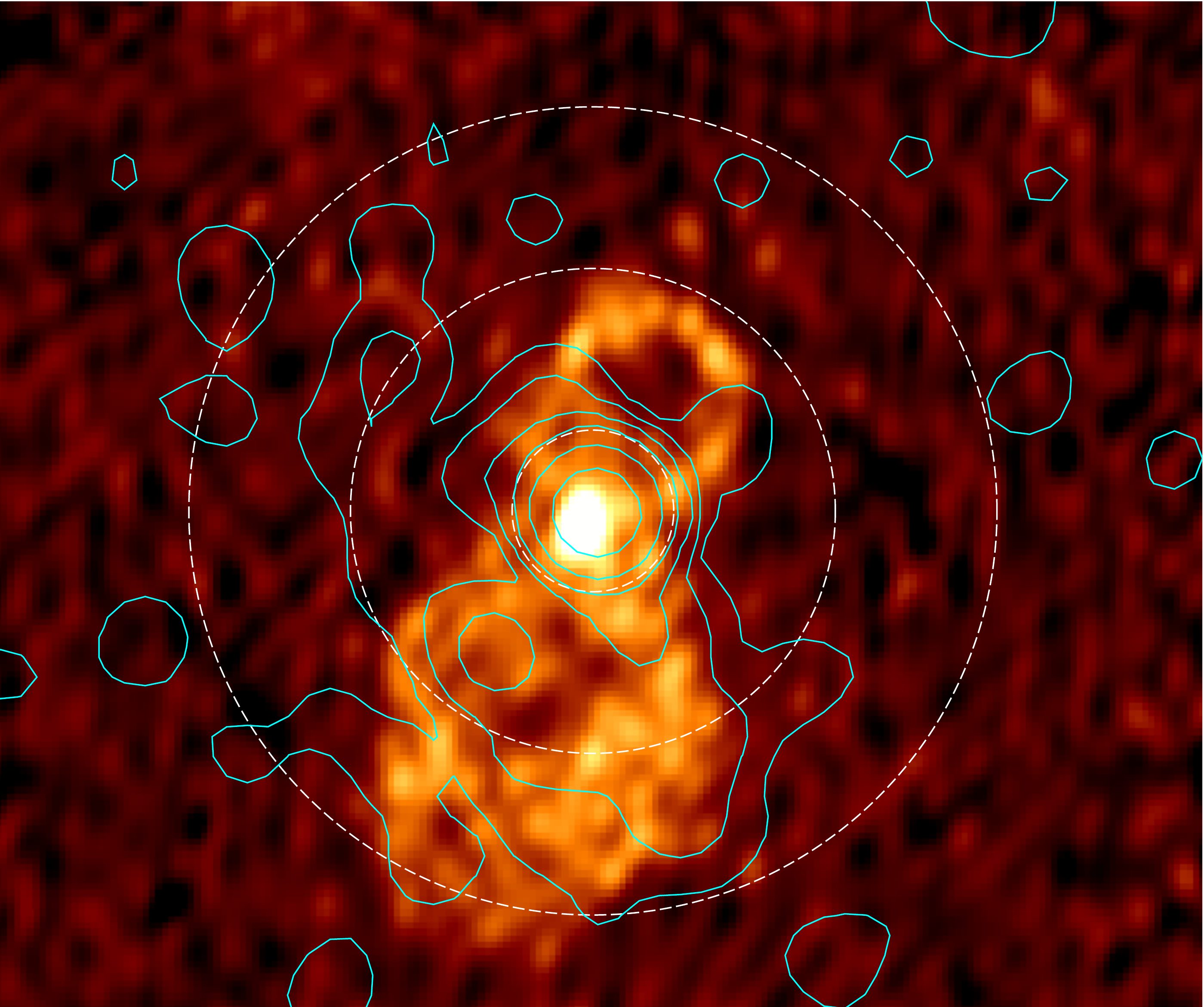}
	\caption{Comparison of the 6 cm VLA figure-eight-shaped structure and the soft X-ray contours (same as Figure~\ref{fig-vband-sx}). 
	The dashed circles are the same as in Figure \ref{fig-vband-sx}\label{fig-radio-sx}}
\end{figure}

\cite{2005ApJ...628..113C} previously found southeast soft X-ray emission spatially superposed over the southeast radio bubble and there exists no bright optical line or continuum emission there. 
They suggested that the extended X-ray emission from the southeast source can be produced by collisional excitation/ionization from outflowing gas which could be due to starburst or AGN.  In Figure \ref{fig-radio-sx}, the main part of the resolved emission is consistent with the southeast part of the radio bubble, especially in the southern region the non-lowest level contours extend along with the bubble, which is more significant than \cite{2005ApJ...628..113C}. It is plausible that the resolved extended soft X-ray emission near the nucleus of NGC 2992 could be dominated by the hot gas heated by shocks from outflows associated with the radio bubble. 
\cite{2010A&A...519A..79F} further concluded that the outflows of NGC 2992 were driven by the AGN rather than the starburst, based on the energy and timescale of starburst that is inconsistent with outflows by using the optical and infrared data. 
In the X-ray band, we also attempt to differentiate the origin of outflows by estimating the expected luminosity of the extended soft X-ray emission, assuming that the soft emission is originated from outflows of starburst. 
From the \textit{apec} model, we calculate an electron density of $n_{e} \approx 3\times10^{-5}\rm\ cm^{-3}$, assuming the volume of a cone with a height of 200 pc, a radius of 200 pc, and $n_{e} \approx n_{H}$. 
\cite{2010A&A...519A..79F} have estimated the mechanical luminosity from stellar winds and SNe is about $10^{40}\rm\ erg\ s^{-1}$. 
Following equation (7) in \cite{1996ApJ...457..616H}, we estimate the expected luminosity of the extended soft X-ray to be $L_{X} \approx 10^{37} - 10^{38}\rm\ erg\ s^{-1}$ \citep[see also][]{2014ApJ...788...54G}, which is $\sim 2$ orders of magnitude lower than the observed result.  
This implies that the extended soft X-ray emission could not be dominated by outflows powered by the starburst.

The radio morphology of NGC 2992 closely resembles the ``Teacup'' AGN \citep{2012MNRAS.420..878K}, a radio-quiet type 2 quasar with ``bubbles'' of radio emission extended to $\sim$10--12 kpc \citep{2015ApJ...800...45H}. The large-scale radio bubbles are thought to be inflated by the AGN, and the outflow is expected to be interacting with the gas on $\sim$10 kpc scales. With {\em Chandra} imaging data, \citet{2018ApJ...856L...1L} reported a 10 kpc loop of X-ray emission, spatially coincident with the luminous eastern radio bubble and [O III] emission line gas. 
Similar to NGC 2992, the extended X-ray is well modeled with a shocked thermal gas with temperature $T = 4-8\times10^{6}$ K \citep{2018ApJ...856L...1L}.
A hotter component with T$\geqslant 3 \times 10^7$ K is possibly also present. In a broader context, there are several well known local Seyfert galaxies with remarkable radio bubbles/lobes, which shows similar characteristic X-ray-emitting gas with a few million K temperatures \citep[e.g.,][]{2010ApJ...719L.208W,2011ApJ...731...21M,2012ApJ...756...39P}. The origin of the extended hot gas is thought to be from interaction between AGN and host ISM, which is shock-heated by AGN-driven outflows.

AGN photoionization could also contribute a significant portion of the extended soft X-ray emission. 
\cite{2017ApJ...840..120M} derived a relatively high G value of Si~{\sc{xiii}}, which indicated that the gas is more likely photoionized \citep{2010SSRv..157..103P}. 
According to our modeling with photoionized component (Figure \ref{fig-spec-cloudy} and Table \ref{table-cloudy}), the soft excess in our spectrum can be well fitted, except the column density of illuminated cloud $\rm log N_{H}$, which is not be well constrained.
Furthermore, the morphology of the extended soft emission is in agreement with the ionization cone in V band image (Figure \ref{fig-vband-sx}) and [O~{\sc{iii}}]$\lambda5007$ \citep{2001A&A...378..787G}, which is suggestive of contribution from photoionized gas.

Taking into account of the northwest region that is obscured by the dust lane shown in the HST image (see Figure \ref{fig-vband-sx}), the total luminosity of the extended soft X-ray emission approaches the prediction of \cite{2005ApJ...628..113C}, $\sim 10^{40}\rm\ erg\ s^{-1}$. In general, the soft excess of NGC 2992 could originate from both AGN-photoionization and collisional excitation/ionization from AGN driven outflow. However, the quality of current spectrum is too poor to fit with both components. 

\subsection{Origin of the 4.97 keV Emission Line} \label{subsec:4.9keV}
The narrow 4.97 keV emission line is unusual in astronomical observation. 
Some other works also showed the presence of narrow emission features at unusual energies \citep[e.g.][]{2011MNRAS.414.1965L,2016MNRAS.455L..26X} and they suggested that these emission lines can probably be generated from the nuclear spallation of Fe by low energy cosmic rays from AGN. 
The process of nuclear spallation can enhance the abundance of sub-Fe elements such as Ti, V, Cr, and Mn, then some neutral emission lines like V $\rm K\alpha$ (4.95 keV), Cr $\rm K\alpha$ (5.41 keV), and Mn $\rm K\alpha$ (5.89 keV) will become strong enough to be detected, and if, in ionized material, there may appear highly ionized emission lines in the X-ray spectrum like Ti~{\sc{xxii}} (4.97 keV), V~{\sc{xxiii}} (5.44 keV), et al \citep{1997ApJ...478..522S,2010ApJ...709.1230T,2019MNRAS.484.3036G}. 
Here, the centroid energy of the anomalous emission line we found is 4.97$^{+0.02}_{-0.02}$ keV, which is consistent with V $\rm K\alpha$ or Ti~{\sc{xxii}} after considering the redshift of NGC 2992 and the error range. 
If considering the 4.97 keV emission line is V $\rm K\alpha$, we can roughly estimate the abundance ratio of V and Fe according to the following relationship: $F_{1}/F_{2} \propto Z_{1}/Z_{2}$, where F is the fluorescence flux of certain elements and Z is the abundance. 
Following \cite{2019MNRAS.484.3036G}, the emission line flux ratio of V and Fe (i.e. $\rm F_{1}/F_{2}$) is about 0.2 with the altering abundances of V and Fe by the effects of spallation, and the abundance of Fe is about half of the solar value from \cite{1989GeCoA..53..197A} in their work. 
Then we estimate the flux ratio of 4.97 keV line and Fe $\rm K\alpha$ is $\sim$ 0.5, assuming $Z_{\rm Fe}=0.5Z_{\rm Fe,\astrosun}$, where  $Z_{\rm Fe,\astrosun}$ is the the solar abundance of Fe.  Considering the rough consistency of our estimation and the result, nuclear spallation could be a plausible mechanism to yield the 4.97 keV emission line. 

Another possible interpretation is that the 4.97 keV may resemble a redshifted emission line like Fe~{\sc{xxv}}, given that the south part of figure-8-shape radio bubbles is in fact redshifted \citep{2001A&A...378..787G,2010A&A...519A..79F}. 
If we consider the emission line is associated with redshifted Fe~{\sc{xxv}} emission ($\rm \sim6.7\ keV$), the projected outflow velocity of this 4.97 keV line is up to $0.23\,c$. If we attribute this emission line to be redshifted $\rm Fe\ K\alpha$, the required velocity is still as high as $0.2\,c$. Only the UFO and the relativistic jet may have such a characteristic high outflow velocity.
Considering the spatial scale much larger than accretion disk and the high velocity of the detected 4.97 keV emission line, it could be produced by a cloud associated with a redshifted jet. If the outflow is bipolar, there may exist a blueshifted high velocity counterpart as well and the expected line centroid will be at $\sim$7.7 keV and $\sim$8 keV for $\rm Fe\ K\alpha$ and Fe~{\sc{xxv}}, respectively. Although previous radio observations reveal presence of radio bubbles and lobes \citep{1984ApJ...285..439U,2017MNRAS.464.1333I}, no other emission/absorption lines related to jet can be found and evidence is lacking for this hypothesis. 

Recently, \citet{2020MNRAS.496.3412M} found an variable emission structure with a possible modulation with a period $T=41\pm 1\rm\,ks$ in the 5.0--5.8 keV band using {\em XMM-Newton} and {\em NuSTAR} data, which is roughly consistent with the energy centroid of the 4.97 keV emission line.
They interpreted this feature as a redshifted Fe K line originated from a emitting region near to the SMBH with a radial distance about $\sim 12\,r_{g}$.
Later \citet{2022MNRAS.514.2974M} also found a variable feature around 5 keV using the latest {\em XMM-Newton} and {\em NuSTAR} observations, and they suggest that this feature could possibly be associated with the result of \citet{2020MNRAS.496.3412M}.
In our nuclear spectrum ($\leqslant 1\arcsec$), the 4.97 keV emission line cannot be detected, which is overwhelmed by the high level of continuum in the nuclear region. Nevertheless, we checked the light curve of the nucleus ($\leqslant 1\arcsec$) in the 4.5--5.5 keV energy band for any modulation, but no statistically significant variability could be found. The lack of modulation signal may well be due to the relatively short coverage of the observations here.

\subsection{Extended Fe $K\alpha$ Emission} \label{subsec:Fe K}

We found the spatially extended ($\sim200$ pc) Fe $\rm K\alpha$ emission along with the disk of the galaxy with a large equivalent width (EW; $\sim2.93^{+6.39}_{-0.50}$ keV) in the Compton thin Seyfert galaxy NGC 2992. 
Normally, the extended Fe $\rm K\alpha$ line emission is expected to arise from reflection of the X-ray continuum by cold and distant material like in other Seyfert 2 galaxies, such as NGC 1068 \citep{2001ApJ...556....6Y, 2015ApJ...812..116B}, NGC 4388 \citep{2021ApJ...908..156Y}, NGC 4945 \citep{2017MNRAS.470.4039M}, and ESO 428-G014 \citep{2017ApJ...842L...4F}. These galaxies contain a Compton thick nucleus with the column density $N_{\rm H} > 1.25\times 10^{24}\rm\ cm^{-2}$ (the inverse of the Thomson scattering cross-section), except NGC 4388 with a column density several times less than the critical value \citep{2021ApJ...908..156Y}. 

The measured line-of-sight column density towards NGC 2992 is $N_{\rm H}= 7.8\pm 0.1 \times 10^{21}\rm\ cm^{-2}$ \citep[][]{2022MNRAS.514.2974M}, even lower than NGC 4388. The reflected emission is likely originated from material with the column density $N_{\rm H}=9.6\pm 2.7\times 10^{22}\rm\ cm^{-2}$\citep{2022MNRAS.514.2974M}.
In Compton thick AGNs, the heavy obscuration of the primary radiation from the nucleus, at least up to 10 keV, allows reflection spectral components from both cold and ionized circumnuclear material including the Fe $\rm K\alpha$ emission can be detected clearly. {Although NGC 2992 is not a Compton thick AGN, the low state of X-ray flux, the deduction of contribution of PSF from the nucleus in the region and the spatially resolved circumnuclear emission may have led to the observed extended Fe $\rm K\alpha$ and the large EW in the extended region where the continuum is very weak.}

It is intriguing to note that there also exist diffuse Fe K$\alpha$ emission in the circumnuclear region of the Galaxy, which spatially corresponds to the distribution of several dense molecular clouds with the column density $N_{\rm H}\sim 1-10\times10^{22}\rm\,cm^{-2}$ \citep[e.g.,][]{2010ApJ...714..732P}.
The EW ($0.7-1\rm\,keV$) of the diffuse Fe K$\alpha$ emission in the various clouds \citep[][]{2010ApJ...714..732P} is lower but comparable with our measurement in the circumnuclear region of NGC 2992. 
This result overall is consistent with the scene that the extended Fe K$\alpha$ emission in NGC 2992 is originated from the reflection of dense molecular gas in the circumnuclear region, which worth further follow up observations in the sub-mm and in the X-rays.

\section{Conclusions} \label{sec:conclusions}

The {\em Chandra}-HETG data of NGC 2992 observed in 2010 are used to investigate the X-ray emission of the circumnuclear region.
Thanks to the historically low state of the AGN in NGC 2992, we overcome the pile-up effect and present spatial and spectroscopy analyses of the circumnuclear region (1\arcsec--3\arcsec).  The main results and conclusions are as follows:

\begin{enumerate}

\item The asymmetric extended soft emission is found with $kT = 0.33\rm\ keV$ and 0.3--2 keV luminosity $L_x \sim 7\times10^{39}\rm\ erg\ s^{-1}$ (Tab.~\ref{table-apec} and Sect.~\ref{subsec:Spectroscopy}).
According to the estimated luminosity of extended soft X-ray emission from starburst is $\sim 2$ orders of magnitude lower than the observed result, the soft excess could not be originated from wind powered by starburst.
From our results, we suggest that the extended soft emission could be produced by both AGN-photoionization and collisional ionization/excitation from outflows produced by AGN.
The majority is originated from the AGN-driven outflows for the consistence between the main part of soft emission and the southeast part of radio figure-8-shape bubbles (Sect.~\ref{subsec:soft excess}).

\item An unusual 4.97 keV emission line has been detected in the southwest region extending to $\sim200$ pc from the nucleus, which may be originated from nuclear spallation of Fe by low energy cosmic rays from AGN. 
Another possible explanation is that the 4.97 keV emission line is a redshifted Fe~{\sc{xxv}} (or Fe K$\alpha$).
Under this situation, the radial velocity of this emission line can be up to $0.23\,c$ (or $0.2\,c$, for redshifted Fe K$\alpha$), implying that this 4.97 keV line may originate from the material associated with a relativistic outflow/jet (see Section~\ref{subsec:4.9keV}).

\item The extended Fe K$\alpha$ emission is also found in the same region, which is a rare feature in Compton thin AGNs.
We suggest that thanks to the historically low state of AGN, the low count rate of zeroth-order image of HETG data and the reduction of PSF contribution from nucleus help reveal the presence of extended Fe K$\alpha$ emission in the spectrum (see Section~\ref{subsec:Fe K}). In analog, there is great chance to capture such extended feature in other highly variable AGN and faded quasars.

\end{enumerate}

More multi-band spatially resolved observations are necessary for investigating the outflows and unique emissions in the circumnuclear region of NGC 2992. For instance, observations using Atacama Large Millimeter/submillimeter Array (ALMA) might help us to understand the nature of outflows from the nucleus of this source if detecting the molecular outflow around the AGN. 
In the future, the next-generation X-ray astronomical satellites like {\em Athena} equipped with revolutionary spectrometer the X-ray Integral Field Unit could provide more information about the X-ray emission near the nuclear regions in AGNs, including the 4.97 keV unique line of NGC 2992.

\acknowledgments

We thank the anonymous referee for very constructive comments that significantly improved our work. We are grateful to Dr. Niel Brandt, Junxian Wang and Xinwen Shu for helpful discussion, and Dr. Fangzheng Shi for technical assistance in line simulations.  J.W. acknowledges support by the National Science Foundation of China (NSFC) grants U1831205 and 12033004, and the science research grants from CMS-CSST-2021-A06 and CMS-CSST-2021-B02. This research has made use of data obtained from the {\em Chandra} Data Archive, and software provided by the {\em Chandra} X-ray Center (CXC) in the application packages CIAO.

\facilities{HST (WFPC2), CXO (HETGS), VLA}

\software{Cloudy \citep{2017RMxAA..53..385F}, 
	CIAO \citep{2006SPIE.6270E..1VF}, DS9 \citep{2003ASPC..295..489J}, ChaRT \citep{2003ASPC..295..477C}, MARX \citep{2012SPIE.8443E..1AD}	
		}

\bibliography{2992}{}
\bibliographystyle{aasjournal}



\end{CJK*}
\end{document}